\documentclass[10pt,aps,prd,letterpaper,typeset,twocolumn,superscriptaddress,showpacs,floatfix]{revtex4-2}
\usepackage{color}
\usepackage{graphicx}
\usepackage{amsmath}
\usepackage{amssymb,amsthm}
\usepackage{hyperref}
\usepackage{mathtools}
\usepackage{physics}
\usepackage{multirow}
\usepackage{caption, subcaption}
 
\graphicspath{{pict/}{}}
\usepackage[applemac]{inputenc}
\usepackage{pgf,tikz}
\usepackage{mathrsfs}
\usepackage{bm}
\usepackage{array}
\usepackage{blkarray}
\usepackage{xcolor}
\usepackage{float}
\usepackage{hyperref}
\hypersetup{colorlinks=true,linkcolor=blue,citecolor=blue,urlcolor=blue}

\begin{document}
\title{The role of Wigner rotation in estimating the specific angular momentum of a Kerr spacetime}

\author{F. J. Lobo}
\affiliation{Facultad de Ciencias F\'isicas y Matem\'aticas, Departamento de F\'isica, Universidad de Concepci\'on, Concepci\'on, Chile}
\affiliation{Instituto Milenio de Investigaci\'on en \'Optica, Universidad de Concepci\'on, Concepci\'on, Chile}
\author{M. Rivera-Tapia}
\affiliation{Facultad de Ciencias F\'isicas y Matem\'aticas, Departamento de F\'isica, Universidad de Concepci\'on, Concepci\'on, Chile}
\affiliation{Instituto Milenio de Investigaci\'on en \'Optica, Universidad de Concepci\'on, Concepci\'on, Chile}
\author{G. Rubilar}
\affiliation{Facultad de Ciencias F\'isicas y Matem\'aticas, Departamento de F\'isica, Universidad de Concepci\'on, Concepci\'on, Chile}
\author{A. Delgado}
\affiliation{Facultad de Ciencias F\'isicas y Matem\'aticas, Departamento de F\'isica, Universidad de Concepci\'on, Concepci\'on, Chile}
\affiliation{Instituto Milenio de Investigaci\'on en \'Optica, Universidad de Concepci\'on, Concepci\'on, Chile}

\begin{abstract}
We study the rotation of the polarization due to the gravitational field in the Kerr spacetime and the possibility of estimating the specific angular momentum that parameterizes this metric. Our approach is based on a geodesic interferometer, that is, a Mach-Zehnder interferometer whose arms are defined by null geodesics, and a single photon propagating within it. We show that the detection probability at the output ports of the interferometer is a function of two phase differences, one arising from the gravitational time delay and the other from the polarization rotation, both computed under the slow rotation and weak field approximations. Thereby, the interferometric visibility is a signature of two relativistic effects. Using the detection probability, we obtain an estimate for the specific angular momentum and characterize its uncertainty.
\end{abstract}

\maketitle

\section{INTRODUCTION}

Reconciling Quantum Mechanics and General Relativity remains a central challenge in contemporary physics \cite{Peres_2004,BELENCHIA_2022}. An important role is played by understanding the interaction between quantum systems and gravitational fields, which may contribute to advancing the development of a unified theory of quantum gravity and to investigating new regimes in fundamental physics. Applied aspects have also gained significant prominence due to advancements in the preparation, control and measurement of quantum systems, which have been instrumental in enhancing the precision of frequency standards \cite{Takamoto_2005} and in measuring local gravitational acceleration with greater accuracy \cite{Poli_2011}. 

Photons are ideal systems for studying the interplay between quantum and gravitational phenomena. This is due to their propagation along null geodesics, which are intrinsically shaped by spacetime geometry \cite{Sereno_2004,Kopeikin_2006}. However, describing a photon as a localized quantum system in curved spacetime presents significant theoretical challenges, stemming in part from the inherent difficulty of achieving strict particle localization within quantum field theory \cite{Brodutch2011}. Nevertheless, for specific applications, a photon can be effectively modeled as a localized qubit, that is, a two-level quantum system defined by its two-dimensional polarization, that is spatially well-localized and associated with a point in spacetime \cite{Brodutch2011}. This description is based on the geometric optics approximation, which allows the analysis of electromagnetic wave propagation in curved spacetime via rays that trace null geodesics \cite{Sereno_2004,Kopeikin2001}. Within this framework, the quantum state of the photon is encoded in its complex polarization vector field, whose evolution is governed by parallel transport along its trajectory \cite{Sereno_2004, Dahal2021}. The applicability of this localized qubit representation typically relies on additional assumptions, such as the confinement of the photon's wave packet to a region significantly smaller than the characteristic scale of spacetime curvature and the effective ``freezing" of its spatial degrees of freedom during transport \cite{Brodutch2011}.

The influence of curved spacetime, particularly in the presence of a rotating gravitational source, manifests itself in the photon internal degree of freedom. Relativistic effects such as the Wigner rotation and gravity-induced polarization rotation - the latter also known as the Skrotskii effect or the gravitational Faraday effect - modify the quantum state of the photon's polarization along its propagation \cite{Noh2021, Basso2021b, Sereno_2004, Kopeikin2001}. The Skrotskii effect, in particular, arises from the gravitomagnetic aspects of the gravitational field generated by a rotating body, analogous to the electromagnetic field of a rotating charged body \cite{Sereno_2004}. Theoretically, the magnitude of this polarization rotation is predicted to depend on the projection of the rotating body's total angular momentum along the photon's line of sight \cite{Sereno_2004}. Although direct experimental measurements of such polarization modifications along astrophysical light paths remain limited \cite{Sereno_2004}, a thorough understanding of these effects is vital for high-precision applications, including astrometry \cite{Kopeikin2001} and satellite-based quantum communications \cite{Noh2021, Silvestri2024, Silvestri2023}.

These relativistic phenomena motivate experimental investigations designed to test fundamental aspects of general relativity and to explore the utility of quantum techniques in precision measurements \cite{Basso2021a, Brodutch2011, Silvestri2024, Silvestri2023}. Interferometers, especially Sagnac and ring laser configurations, are powerful experimental tools for quantifying rotational effects \cite{Delgado_2002,Kajari_2004,Feiler_2009}, including those due to Earth's rotation \cite{Ullah2025, Schreiber2023, Lai2020}. Recent experiments have successfully measured the phase difference induced by Earth's rotation on entangled two-photon states using a large-scale fiber Sagnac interferometer \cite{Silvestri2024, Silvestri2023}, thereby demonstrating the capacity of quantum information resources, such as entanglement, to enhance sensitivity in such measurements \cite{Brodutch2011, Basso2021b}. A direct measurement of the Skrotskii effect, for instance, would constitute a significant test of inertial frame dragging \cite{Sereno_2004}, a general relativistic effect predicted to occur in the vicinity of massive rotating bodies, such as Kerr black holes.

In this article, we study the rotation of the polarization of a photon in the Kerr spacetime metric and the possibility of estimating from this the specific angular momentum that parameterizes the Kerr metric. In particular, we define a Mach-Zehnder geodesic interferometer, i.e., an interferometer whose arms correspond to null geodesics. This result is obtained without any approximations. We characterize the phase differences between both arms generated by the arrival time difference and the polarization rotation, which are obtained under the slow rotation and weak field approximations. For a geodesic interferometer of approximately 1 ${\rm m}^2$ of area situated close to the Earth's surface and light of frecuency $10^{12}$ Hz the phase difference generated by the arrival time delay is of the order of $10^{-16}$ .
For similar conditions, the Wigner phase difference is of the order of $10^{-30}$. This value can be increased to $10^{-23}$  by increasing the distance between the mirrors of the geodesic interferometer to $\sim 10^3$ m. We also calculate the quantum state of a single photon after propagating inside the interferometer and obtain the probability of detecting the photon at each of the output ports of the interferometer. This probability and the characteristic interferometric visibility are functions of both phase differences. This result is then used to obtain an estimate of the specific angular momentum in terms of the detection probability and also obtain an expression for the uncertainty of the estimate. For a geodesic interferometer close to Earth whose mirrors are separated by a radial coordinate interval of the order of 800 km and an error in the probability of $10^{-14}$, the relative error in the estimation of the specific angular momentum is in the order of $10^{-6}$.

\section{Preliminars}
\label{Preliminars}

In this section, we introduce the Kerr spacetime metric and its corresponding null geodesics. Thereafter, we introduce the tetrad formalism to analyze the polarization rotation of a photon along a null geodesic within the Wentzel-Kramers-Brillouin (WKB) approximation. In particular, we derive two coupled partial differential equations that govern the polarization rotation. Afterward, we connect with the polarization of the quantized electromagnetic field.

\subsection{Kerr metric}
\label{Kerr metric}

 The Kerr metric describes the spacetime geometry generated by a rotating black hole of mass $M$ and angular momentum $J$.  
In Boyer-Lindquist coordinates $x^\mu = (ct, r, \theta, \varphi)$, the line element is given by $ds^2 = g_{\mu\nu}dx^\mu dx^\nu$, where the non-zero components of the metric tensor are defined as follows:
\begin{subequations}\label{eq:KerrComponents}
\begin{align}
    g_{00} &= \left(1 - \frac{2mr}{\rho^2}\right), \label{eq:g_tt} \\
    g_{11} &= -\frac{\rho^2}{\Delta}, \label{eq:g_rr} \\
    g_{22} &= -\rho^2, \label{eq:g_thetatheta} \\
    g_{33} &= -\left(r^2+a^2 + \frac{2mra^2\sin^2\theta}{\rho^2}\right)\sin^2\theta, \label{eq:g_phiphi} \\
    g_{03} &= g_{30} = -\frac{2mar\sin^2\theta}{\rho^2}. \label{eq:g_tphi}
\end{align}
\end{subequations}
In the expressions above, $m \equiv GM/c^2$ represents the geometric mass of the black hole, $a \equiv J/(Mc)$ is the specific angular momentum, and the functions $\rho^2$ and $\Delta$ are defined as
\begin{align}
    \rho^2 &\equiv r^2 + a^2\cos^2\theta, \\
    \Delta &\equiv r^2 - 2mr + a^2.
\end{align}
In the far-field limit ($r \gg m, a$), the Kerr metric reduces to the weak-field solution for a rotating mass source \cite{Mashhoon2003}. A comparison of the metric elements in Eqs.~\eqref{eq:g_tt}-\eqref{eq:g_tphi} with solutions derived from linearized gravity \cite{1995grin.book.....C}, confirms the identification of $M$ as the total mass and $J$ as the total angular momentum of the source.

\subsubsection{Principal null geodesics}\label{section_geodesicas_kerr}

Although the Schwarzschild spacetime admits purely radial null geodesics, the Kerr spacetime lacks the spherical symmetry required for such trajectories \cite{MTW1973}. For the Kerr metric, we derive general analytical solutions for null geodesics on arbitrary trajectories (see Appendix \ref{Appendix:General_nullgeodesics} for details). However, for simplicity, in this work we focus on the special case of principal null geodesics, which are confined to a cone of constant polar angle, $\theta = \text{const}$.

We parameterize null geodesics using an affine parameter $\lambda$, such that the null tangent vector is $k^{\mu} \equiv dx^{\mu}/d\lambda$. For the principal null geodesics under consideration ($\theta = \text{const}$), the radial component $k^r = dr/d\lambda$ is constant along the path. We are free to rescale $\lambda$ to set this constant, thereby adopting the parameterization $dr/d\lambda = \epsilon_r = \pm 1$ with $\epsilon_r=+1$ for outgoing and $\epsilon_r=-1$ for ingoing principal null geodesics. Thus, the geodesic equations for the remaining coordinates $t$ and $\varphi$ become (see Appendix \ref{Appendix:kerr_geodesic} for details)
\begin{equation}
\label{eq:kerr_geodesics_diff}
    c\frac{dt}{dr} = \epsilon_r\frac{r^{2} + a^{2}}{\Delta}~~~{\rm and}~~~\frac{d\varphi}{dr} = \epsilon_r\frac{a}{\Delta},
\end{equation}
respectively.

Integrating these equations yields the trajectory solutions. The solution in the case $a>m$ for the azimuthal angle $\varphi(r)$, starting from a radius $r_0$ with angle $\varphi_0$, is
\begin{align}\label{eq:phi_sol}
    \varphi(r) &= \varphi_0 +\epsilon_r\frac{a}{\sqrt{a^2 -m^2}}\left(\arctan\left( \frac{r-m}{\sqrt{a^2 - m^2}}\right)\right.  \nonumber\\
    & \left. - \arctan\left(\frac{r_0 - m}{\sqrt{a^2 - m^2}} \right)\right),
\end{align}
and the solution for the time coordinate $t(r)$ is
\begin{align}
    ct(r) &= ct_0 + \epsilon_r(r-r_0)\nonumber \\
    & +\epsilon_r\frac{2m^2}{\sqrt{a^2 - m^2}}\left(\arctan\left( \frac{r-m}{\sqrt{a^2 - m^2}}\right) \right.\\
    &-\left. \arctan\left(\frac{r_0 - m}{\sqrt{a^2 - m^2}} \right) \right) + \epsilon_r m \ln\left| \frac{r^2 -2mr +a^2}{r_0^2 -2mr_0 + a^2}\right|,\nonumber
\end{align}
are the radii of the outer and inner event horizons, respectively.

The null tangent vector, parameterized by the radial coordinate $r$, is $u^{\mu} \equiv dx^{\mu}/dr = (c(dt/dr), dr/dr, d\theta/dr, d\varphi/dr)$. Using the differential relations from Eq.~\eqref{eq:kerr_geodesics_diff} and the fact that $\theta$ is constant, the contravariant components of this vector are
\begin{equation}\label{eq:null_vector}
    u^{\mu} = \left(\epsilon_r \frac{r^2+a^2}{\Delta}, 1, 0, \epsilon_r \frac{a}{\Delta}\right).
\end{equation}


\subsection{Tetrads and local reference frames}
\label{Tetrad-local}

To analyze the polarization of a photon propagating in Kerr spacetime, we construct an orthonormal basis $\{e^\mu_{\hat{a}}\}$ of four vectors, known as a tetrad, at each point along the photon's path \cite{PALMER20121078}. This basis defines a local  reference frame, where the basis vectors satisfy the relation $g_{\mu\nu}e^\mu_{\hat{a}}e^\nu_{\hat{b}} = \eta_{\hat{a}\hat{b}}$, with $\eta_{\hat{a}\hat{b}} = \text{diag}(1, -1, -1, -1)$.

We align this frame with the principal null geodesic derived in the previous section. The null tangent vector $u^\mu$ from Eq.~\eqref{eq:null_vector} is used to construct a null tetrad. First, we define a rescaled null tangent vector $\bar{u}^\mu = u^\mu / N$, which is then expressed as a combination of the timelike and one of the spacelike basis vectors, that is,
\begin{equation}\label{eq:null_tetrad_relation}
    \bar{u}^{\mu} = e^{\mu}_{\hat{0}} + e^{\mu}_{\hat{3}}.
\end{equation}
The scaling factor $N$ is fixed by the normalization condition that the projection of $\bar{u}^\mu$ onto the local time axis is unity, which using the dual basis $e^{\hat{0}}_{\mu}$ is given by
\begin{equation}\label{eq:rescaling_condition}
    e^{\hat{0}}_{\mu} \bar{u}^{\mu} = 1.
\end{equation}
This construction ensures that the vector $e^\mu_{\hat{3}}$ is aligned with the spatial direction of the photon propagation in the local frame. Following this scaling, the components of the null vector on the basis of tetrads, denoted $\bar{u}^{I}$, take the canonical form:
\begin{equation}
    \bar{u}^{\hat{a}} = \left(1,0,0,1\right).
\end{equation}

To find an explicit form for the tetrad, we solve the orthonormality conditions $g_{\mu\nu}e^\mu_{\hat{a}}e^\nu_{\hat{b}} = \eta_{\hat{a}\hat{b}}$, adopting the signature $\eta_{\hat{a}\hat{b}} = \text{diag}(1, -1, -1, -1)$. Our strategy is to choose the basis vectors by making physically motivated choices. First, we fix the timelike vector of the tetrad by aligning $e^\mu_{\hat{0}}$ with the four-velocity of a static observer \cite{Semerak1993,Kohlrus2018}. This is well-defined in the region outside the ergosphere where the time-translation Killing vector $\partial_t$ is timelike ($g_{00} > 0$), that is,
\begin{equation}\label{eq:e0_def}
    e^\mu_{\hat{0}} = \frac{1}{\sqrt{g_{00}}} \partial_t = \left(\frac{1}{\sqrt{g_{00}}}, 0, 0, 0\right).
\end{equation}
Next, using the rescaled null vector $\bar{u}^\mu$ from Eq.~\eqref{eq:null_tetrad_relation}, we define the basis vector along the direction of propagation. To maintain consistency with the relation~\eqref{eq:null_tetrad_relation}, we must have that
\begin{equation}\label{eq:e3_def}
    e^\mu_{\hat{3}} = \bar{u}^{\mu} - e^{\mu}_{\hat{0}}.
\end{equation}
These choices for $e^\mu_{\hat{0}}$ and $e^\mu_{\hat{3}}$ significantly constrain the system. The remaining freedom lies in the choice of the two transverse basis vectors, $e^\mu_{\hat{1}}$ and $e^\mu_{\hat{2}}$, which correspond to a rotation in the plane orthogonal to the direction of propagation. This leaves a single free parameter, a rotation angle, to fully specify the tetrad.

In the Schwarzschild limit, four of the solutions containing a free component diverge and are therefore discarded. The four remaining viable solutions differ only by an overall sign. As the physical effect must be independent of the chosen tetrad basis, a single representative solution is selected without loss of generality, which is given by

\begin{align}
e^{\mu}_{\hat{0}}&= \left(\frac{1}{\sqrt{g_{00}}},0,0,0\right), \\
e^{\mu}_{\hat{1}}&=\left(\frac{g_{03} \mathcal{U}}{\sqrt{\Lambda \alpha}}, \quad
-\frac{\bar{u}^{1} \bar{u}^{3} \Delta}{\mathcal{U} \sqrt{\Lambda \alpha}}, \quad
0, \quad
-\frac{g_{00} \mathcal{U} }{\sqrt{\Lambda \alpha}}\right), \\
e^{\mu}_{\hat{2}}&= \left(0,0,-\frac{g_{11} \bar{u}^{1}}{\mathcal{U}\sqrt{-g_{22} g_{11} }},0\right), \\
e^{\mu}_{\hat{3}}&= \left(\bar{u}^{0}-\frac{1}{\sqrt{g_{00}}},\bar{u}^{1},0,\bar{u}^{3}\right), \\[1em]
\end{align}
where $\Lambda \equiv g_{00} g_{33}-g_{03}^2$, $\alpha \equiv g_{00} \left[-g_{11} (\bar{u}^{1})^2-g_{33} (\bar{u}^{3})^2\right]+g_{03}^2 (\bar{u}^3)^2$ and $\mathcal{U} \equiv \sqrt{g_{11} (\bar{u}^{1})^2}$. The scaling $N$ of the null vector can be obtained by direct calculation, leading to
\begin{equation}
    N = \frac{g_{00} u^{0} + g_{03} u^{3}}{\sqrt{g_{00}}} = u^{\hat{0}},
\end{equation}

\noindent where $u^{\hat{0}}$ is the zeroth component of the null vector $u^{\mu}$ on the adapted tetrad basis.

\subsection{Polarization rotation in curved spacetime }

Our analysis utilizes the mathematical framework developed in \cite{PALMER20121078}, which has seen subsequent application in various fields, notably quantum field theory in curved spacetimes \cite{PhysRevA.99.032350, Marzlin_2022, Basso2021b}. To investigate phenomena such as Wigner rotation in this context, we employ the WKB approximation. This method involves an ansatz for the vector potential of the elctromagnetic field of the form $A_{\mu} = \operatorname{Re}\left[\varphi_{\mu} e^{i \varrho} \right]$, where $\varphi_{\mu}$ is a slowly-varying complex amplitude and $\varrho$ is a rapidly oscillating phase.

A precise description of the quantum states of light is essential for the present analysis. We accomplish this by further employing the tetrad formalism, introduced in the previous section, to define the null vector $u^{\mu}$. Within this framework, quantum states of light can be identified through two main approaches: using an adapted tetrad or a non-adapted tetrad. The choice between these approaches is fundamental because it dictates how the polarization state is subsequently described and manipulated.

We employ the adapted-tetrad method, which involves aligning the tetrad vector $e^{\mu}_{\hat{0}}+e^{\mu}_{\hat{3}}$ with the null vector $u^{\mu}$ that defines the direction of wave propagation. This choice was made specifically to isolate the polarization degree of freedom of light. The resulting polarization vector on the tetrad basis, $\psi^{\hat{I}}$, can then be expressed in this basis in the form $(\nu, \psi^{\hat{1}}, \psi^{\hat{2}}, \nu)^T$ \cite{PALMER20121078}, since $u_{\hat{I}}\psi^{\hat{I}}=\psi^{\hat{0}}- \psi^{\hat{3}}=0$, then $\psi^{\hat{0}}=\psi^{\hat{3}}=\nu$. Here, the complex amplitudes $\psi^{\hat{1}}$ and $\psi^{\hat{2}}$ correspond to the horizontal and vertical components of the state on a linear polarization basis. Consequently, the relevant two-dimensional quantum state is concisely represented by the Jones vector $\vert\psi\rangle = (\psi^{\hat{1}}, \psi^{\hat{2}})^T$. 

In the context of the transport of a localized photon wave packet, it is possible to derive an equation governing the evolution of the polarization vector under rotations induced by the gravitational field \cite{PALMER20121078}. The primary physical effect of this transport is the acquisition of a polarization-dependent phase that changes the polarization state along the propagation of light.
 
In the following, we describe the method to obtain the Wigner phase. First, the change of the polarization along the propagation of the photon in the adapted-tetrad frame is given by
\begin{eqnarray}
\frac{d\psi^{\hat{A}}}{d\lambda} + W^{\hat{A}}_{\;\hat{B}} \psi^{\hat{B}} &=& 0,  
\label{Eq_Rotation_polarization}
\end{eqnarray}
where $\hat{A},\hat{B}\; \in \lbrace 1,2 \rbrace$ denotes the components in the linear polarization basis and $W^{\hat{A}}_{\;\hat{B}}$ is the Wigner rotation matrix. As we have seen above, the identification of the state will depend on the choice of the tetrads, and this choice leads to different Wigner rotations. Then, for the adapted-tetrad method, we can calculate the Wigner rotation as
\begin{equation}
W^{\hat{A}}_{\;\hat{B}} =iu^{\mu}\omega_{\mu \hat{1}\hat{2}}(\sigma_y)^{\hat{A}}_{\; \hat{B}},   
\end{equation}
where $\omega_{\nu \; \hat{J}}^{\;\hat{I}}$ is the spin-1 connection 
\begin{equation}
\omega_{\nu \; \hat{J}}^{\;\hat{I}} := e^{\hat{I}}_{\sigma} \partial_{\nu}e^{\sigma}_{\hat{J}} + \Gamma^{\sigma}_{\nu \rho} e^{\hat{I}}_{\sigma} e^{\rho}_{\hat{J}}
\label{spin-1}
\end{equation}
and $\Gamma^{\sigma}_{\mu \rho}$ are the Christoffel symbols.

\begin{equation}\label{transportequation_system}
     \frac{d \psi^{\hat{1}}}{d\lambda} - W^{\hat{1}}_{\;\hat{2}}\psi^{\hat{2}} = 0 \qquad {\rm and} \qquad \frac{d \psi^{\hat{2}}}{d\lambda}+W^{\hat{1}}_{\;\hat{2}}\psi^{\hat{1}} = 0,
 \end{equation}
where 
\begin{align}\label{wigner_rotation}
 W^{\hat{1}}_{\;\hat{2}} = u^{\mu}\omega_{\mu \hat{1}\hat{2}} = -u^{\mu}\omega_{\mu \hat{2}\hat{1}}.   
\end{align}

\subsection{Quantum states of light and Wigner rotation}
\label{Quantum states of light and Wigner rotation}

We consider a pure quantum state of a single photon with frequency dispersion~\cite{Rodriguez_2023} of the form
\begin{equation}
    \ket{\psi} = \frac{1}{\sqrt{2}} \sum_{s = \pm 1} \int d\mathbf{k} f(\mathbf{k}) \ket{\mathbf{k},s}, 
\end{equation}
where $f(\mathbf{k})$ describes the frecuency dispersion, $\ket{ \mathbf{k},s}$ are momentum-helicity eigenstates \cite{Kohlrus2018} of a single photon with $\mathbf{k} = (k^{0},k^{1},k^{2},k^{3})$, $|s\rangle = (|H\rangle - is|V\rangle)/\sqrt{2}$ with $s=\pm1$, and $|H\rangle$ and $|V\rangle$ are the components in the tetrad frame of the quantum state on the basis of linear polarization. The state $\ket{\psi}$ in the local frame \cite{Kohlrus2018} is given by
\begin{equation}\label{initial_photon_state}
    \ket{\psi} = \frac{1}{\sqrt{2}} \sum_{s = \pm 1} \int d\hat{\mathbf{k}} f(\hat{\mathbf{k}}) \ket{\hat{\mathbf{k}},s}, 
\end{equation}
where $\hat{\mathbf{k}} = (k^{\hat{0}},k^{\hat{1}},k^{\hat{2}},k^{\hat{3}})$ are the null vector components on the basis of the adapted tetrad.

The evolution of a momentum-helicity eigenstate is given by a unitary operator as
\begin{equation}
    U\ket{\hat{\mathbf{k}},s} = e^{is\vartheta(\hat{\textbf{k}})}\ket{\hat{\mathbf{k}}_f,s},
\end{equation}
with $\hat{\mathbf{k}}_f$ the local frame momentum at the end point $\lambda_f$ and $\vartheta(\hat{\textbf{k}})$ the Wigner phase. Each state acquires a different Wigner phase factor given by the helicity $s$ \cite{PALMER20121078}. The Wigner phase is obtained by integrating the Wigner rotation along the photon's null geodesic, that is,
\begin{equation}
\label{Wigner_phase_expresion2}
    \vartheta(\hat{\textbf{k}}) = \int^{\lambda_f}_{\lambda_i} W^{\hat{1}}_{\;\hat{2}} d\lambda.
\end{equation}


\section{Polarization rotation and Wigner phase in Kerr spacetime}

In this section, we obtain analytic expressions for the polarization rotation and the Wigner phase for the case of Kerr spacetime and two static observers. Next, we define the geodesic interferometer, calculate its visibility, and estimate the specific angular momentum parameter $a$ of the Kerr spacetime.

\subsection{Polarization rotation and Wigner phase} 
\label{Wigner rotation}

To calculate the Wigner rotation Eq.~\eqref{wigner_rotation}, we use the null vector Eq.~\eqref{eq:null_vector}, obtaining the expression

\begin{align}
    W^{\hat{A}}_{\;\hat{B}}
     &=i\epsilon_r\frac{(a^2 + r^{2})}{\Delta} \omega_{0 \hat{1}\hat{2}} \left(\sigma_y\right)^{\hat{A}}_{\;B} + i\,\omega_{1 \hat{1}\hat{2}}\left(\sigma_y\right)^{\hat{A}}_{\;\hat{B}}\nonumber\\
     & \quad +i\epsilon_r\frac{a}{\Delta} \,\omega_{3 \hat{1}\hat{2}}\left(\sigma_y\right)^{\hat{A}}_{\;\hat{B}},
\end{align}
where $\omega_{\mu \hat{1}\hat{2}}$ is the spin-1 connection in Eq.~\eqref{spin-1}. This leads to the following expression for the Wigner phase 
\begin{align}\label{Wigner_phaseKerr}
    \vartheta = \int^{r_f}_{r_i}\left[\epsilon_r\frac{(a^2 + r^{2})}{\Delta} \omega_{0 \hat{1}\hat{2}}  +\,\omega_{1 \hat{1}\hat{2}}
     +\epsilon_r\frac{a}{\Delta} \,\omega_{3 \hat{1}\hat{2}}\right]dr.
\end{align}

We now specialize the Wigner phase $\vartheta$ to the case of light traveling between two static observers, both located at a constant polar angle $\theta$ and at initial and final radial positions $r_i$ and $r_f$, respectively. In the weak-field and slow-rotation limits, we expand the argument of Eq.~\eqref{Wigner_phaseKerr} into powers of $(a/r)$ and $(m/r)$. Thereby, the Wigner phase, up to fourth order, is given by

\begin{align}\label{wigner_phase_between_two_observers}
     \vartheta =&\left[ \epsilon_r\frac{a \cos (\theta )}{r  } - \epsilon_r\frac{a^{3}\cos(\theta)}{r^{3}}+\frac{a^{3}\cos(\theta)}{6r^{3}}\right.\nonumber \\
     & \;\left.- \frac{a^{3}\cos(\theta)\cos(2\theta)}{6r^{3}} - \epsilon_r\frac{ma^{3}\cos(\theta)\sin^{2}(\theta)}{2r^{4}}\right.
     \nonumber\\
     & \; \left. + \frac{m a^{3}\cos(\theta)\sin^{2}(\theta)}{2r^{4}} + O\left(\frac{1}{r}\right)^{5}\right]^{r_f}_{r_i}.
\end{align}
This expression is consistent with the results obtained previously up to the first order in $(a/r)$ and $(m/r)$ \cite{Fayos1982}. Note that the Wigner phase vanishes for $\theta= \pi/2$. The maximum accumulated Wigner  phase is achieved for a light pulse emitted from infinity to an observer located at $\theta = 0$. Thus, for Earth-like values of the radius and rotation parameter, $R_{\rm Earth} \approx 7 \times 10^{6} \, \mathrm{m}$ and $a_{\rm Earth} \approx 3.97\, \mathrm{m}$, the value of the Wigner phase is approximately $\vartheta \approx 10^{-6}$.

\subsection{Geodesic interferometer}
\label{Geodesic interferometer}

\begin{figure}[H]
        \centering
    \includegraphics[width=8cm]{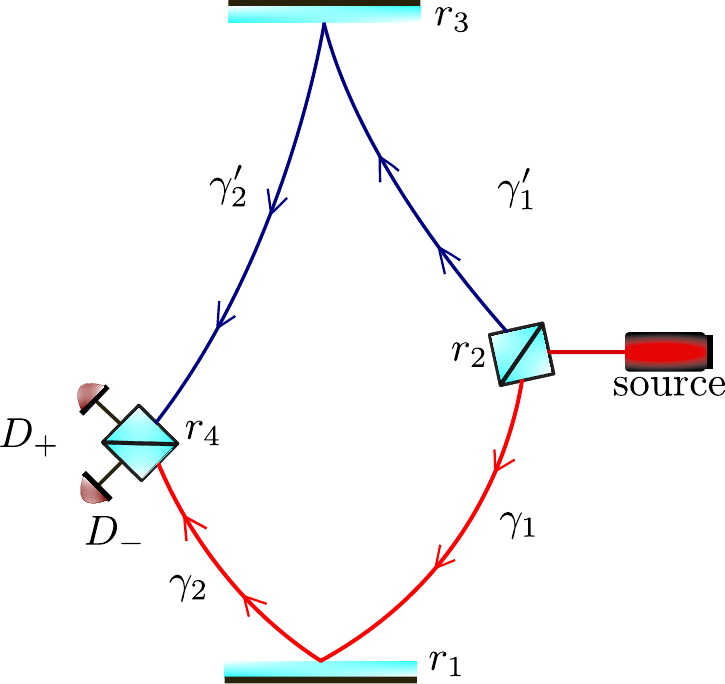}
        \caption{Geodesic interferometer: A single photon emitted from a source at radial coordinate $r_2$ is prepared in a superposition of states propagating along two paths. One path component travels towards $r_1$ and reflects, while the other travels towards $r_3$ and reflects. Both reflected components subsequently propagate to the radial coordinate $r_4$, where they are recombined and measured.}
        \label{fig:Geodesic_Interferometer}
\end{figure}

To investigate relativistic effects on light, we analyze its behavior in free space, specifically considering light that propagate freely along geodesic trajectories. For a single-photon interference experiment, we consider the setup depicted in Fig.\,\ref{fig:Geodesic_Interferometer}. Here, a light source, located at the radial coordinate $r_2$, emits a single photon that interacts with a balanced beamsplitter. Thus, the photon is prepared in a superposition of states propagating along two distinct geodesic paths $\gamma'_1$ and $\gamma_1$. One path extends from coordinate $r_2$ to a lower radial coordinate $r_1$, while the other extends from coordinate $r_2$ to a higher radial coordinate $r_3$. Subsequently, upon reaching the coordinates $r_1$ and $r_3$, the photon undergoes a reflection. The light reflected at coordinate $r_1$ propagates along the geodesic path $\gamma_2$ towards coordinate $r_4$, and  the state reflected at coordinate $r_3$ propagates along the geodesic path $\gamma'_2$ towards coordinate $r_4$. The photon states are then recombined by a balanced beamsplitter at the radial coordinate $r_4$, where the light is finally detected. The radial coordinate $r_4$ satisfies $r_1 < r_4 < r_3$ and its value depends on the other three radial coordinates $r_1, r_2,$ and  $r_3$.

One could consider defining a Mach-Zehnder-type interferometer where trajectories mainly along the radial direction connect two different radial coordinates, $r_1$ and $r_2$, while other segments of the trajectory remain at these fixed radii. This is the case of the Schwarzschild metric because in this spacetime it is possible that light follows purely radial geodesic paths, allowing for such a configuration in free space. However, in the Kerr metric, purely radial geodesic paths generally do not exist because of frame-dragging effects. Consequently, constructing a Mach-Zehnder-type interferometer in this specific manner is feasible in Schwarzschild spacetime but generally not possible in Kerr spacetime.

To determine the coordinate at which light recombines in the geodesic interferometer, the radial coordinate $r_4$ must be calculated. This coordinate is defined by the convergence of photons exiting the beamsplitter at $r_2$ and radially reflected by the mirrors at coordinates $r_1$ and $r_3$, satisfying the condition
$\varphi(r_4, r_0 = r_1)) =  \varphi(r_4, r_0 = r_3)$. 
This calculation uses the solutions for the incoming null geodesic $\varphi_{-}(r)$ and the outgoing null geodesic $\varphi_{+}(r)$. The first path $\gamma_1$ corresponds to the trajectory between the coordinates $r_2$ and $r_1$, given by
\begin{align}
    \varphi_{-}(r_1) &= -\alpha\left(\arctan\left( \frac{r_1-m}{\beta}\right)\right.\nonumber \\
     & -\left.  \arctan\left(\frac{r_2 - m}{\beta} \right) \right),
\end{align}
where we have defined the constants $\alpha$ and $\beta$ as
\begin{equation}
\alpha \equiv \frac{a}{\sqrt{a^2 -m^2}} \; \; {\rm and} \; \; \beta \equiv{\sqrt{a^2 -m^2}}.    
\end{equation}
Analogously, for paths $\gamma_2$, $\gamma'_1$, and  $\gamma'_2$, we have the solutions
\begin{align}\label{varphi_r4_plus}
    \varphi_{+}(r_4) &=  \alpha\left(\arctan\left( \frac{r_4-m}{\beta}\right)\right.\nonumber \\
     & -\left.  \arctan\left(\frac{r_1 - m}{\beta} \right) \right) + \varphi_{-}(r_1),
\end{align}
\begin{align}
    \varphi_{+}(r_3) &= \alpha\left(\arctan\left( \frac{r_3-m}{\beta}\right)\right.\nonumber \\
     & - \left. \arctan\left(\frac{r_2 - m}{\beta} \right)\right) ,
\end{align}
and
\begin{align}\label{varphi_r4_minus}
    \varphi_{-}(r_4) &= -\alpha\left(\arctan\left( \frac{r_4-m}{\beta}\right)\right.\nonumber \\
     & -\left.  \arctan\left(\frac{r_3 - m}{\beta} \right) \right) + \varphi_{+}(r_3),
\end{align}
respectively.

Equating Eqs.~\eqref{varphi_r4_plus} and \eqref{varphi_r4_minus}, that is, $\varphi_+(r_4)=\varphi_{-}(r_4)$, and solving for the radial coordinate $r_4$,  we obtain two solutions, one of them being negative which is discarded. The remaining solution is

This result shows that it is possible to define a Mach-Zehnder interferometer whose arms are given by null geodesics in Kerr spacetime. In the following, we calculate the arrival time difference of photons propagating inside the geodesic interferometer.

For a static observer in the Kerr spacetime, we have the following relationship between coordinate time and proper time
\begin{equation}
    c^{2}d\tau^{2} =\frac{c^{2}dt^{2}}{\rho^{2}}\left( \rho^{2} - 2mr\right),
\end{equation}
which after integration leads to
\begin{equation}
ct =  c \tau\frac{\rho}{\sqrt{\rho^{2} -  2mr}}.
\end{equation}
Thus, for a detector located at the radius $r_{i}$ a static observer has a four-velocity $x^{\mu}_{r_i}(\tau)=\left(ct,r_i, \varphi(r_i),\theta\right)$ given by
\begin{equation}
     x^{\mu}_{r_i}(\tau)= \left(c \tau\frac{\rho}{\sqrt{\rho^{2} -  2mr}},r_i, \varphi(r_i),\theta\right).
\end{equation}
Now, a photon worldline is given by
\begin{eqnarray}
    x^{\mu}_{\rm ph}(\lambda)&=& \left(ct(\lambda),r, \varphi(\lambda),\theta\right),
\end{eqnarray}
where $\lambda$ is the affine parameter. The worldline described by a photon, initially located at  $r_2$, following path $\gamma_1$, will reach the mirror located at $r_1$ having a four-position given by
\begin{equation}
    x^{\mu}_{\rm ph}(r_1) = \left( ct_{-}(r_1,_2),r_1, \varphi_{1}(r_1,r_2),\theta\right).
\end{equation}
This photon reflects radially off the mirror, follows the path $\gamma_2$ and then reaches the detector located at $r_4$ with four-position given by
\begin{eqnarray}
    x^{\mu}_{\rm ph}(r_4) &=& \left(ct_{+}(r_4,r_1) + ct_{-}(r_1,r_2),r_4,\right.\nonumber
    \\
    &&\left. \varphi_{+}(r_4,r_3)+\varphi_{-}(r_1,r_2), \theta\right),
\end{eqnarray}
where we have 
\begin{align}
    ct_{\pm}(r,r_0) &=  \pm \Delta r  \pm m \ln\left| \frac{r^2 -2mr +a^2}{r_0^2 -2mr_0 + a^2}\right|\nonumber \\
    & \pm\frac{2m^2}{\sqrt{a^2 - m^2}}\left(\arctan\left( \frac{r-m}{\beta}\right) \right.\\
    &-\left. \arctan\left(\frac{r_0 - m}{\beta} \right) \right) + ct_0,\nonumber
\end{align}
with  $\Delta r = r -r_0$ and $c t_0$ the initial temporal coordinate. Analogously, for the path $\gamma'_1$, we have the four-position
\begin{eqnarray}
    x^{\mu}_{\rm ph}(r_3) = \left( ct_{+}(r_3,r_2),r_3, \varphi_{+}(r_3,r_2), \theta \right),
\end{eqnarray}
which after reflecting on the mirror at $r_3$ reaches the detector at $r_4$ with four-position given by
\begin{eqnarray}
    x^{\mu}_{\rm ph}(r_4) &=& \left(ct_{-}(r_4,r_3) + ct_{+}(r_3,r_2),r_4,\right. \nonumber
    \\
    &&\left. \varphi_{-}(r_4,r_3)+\varphi_{+}(r_3,r_2),\theta\right).
\end{eqnarray}

To calculate the difference in arrival time, we first obtain the flight time measured by the detector located at $r=r_4$ for each arm of the geodesic interferometer. For the path $(\gamma_1,\gamma_2)$ we have 
\begin{eqnarray}\label{proper_time_noprim}
    c\tau=\Pi_4( ct_{+}(r_4,r_1) + ct_{-}(r_1,r_2))
\end{eqnarray}
while for the path $(\gamma'_1,\gamma'_2)$ we obtain
\begin{eqnarray}\label{proper_time_prim}
    c\tau' =\Pi_4(ct_{-}(r_4,r_3) + ct_{+}(r_3,r_2)),
\end{eqnarray}
with $\Pi_4 = \sqrt{\rho_4^{2} - 2mr_4}/\rho_4$.

Thus, the arrival proper time difference $c\Delta\tau=c(\tau'-\tau)$ between both paths of the interferometer is given by

\begin{align}\label{arrivals_times_difference}
    \frac{c\Delta\tau}{\Pi_4} &= \; 2 \left(r_3 + r_1 -r_4-r_2\right) \nonumber\\
    & - \frac{4m^2}{\sqrt{a^2 -m^2}} \left( \arctan\left( \frac{r_4 - m}{\beta}\right) \right. \nonumber\\ 
    &- \arctan\left( \frac{r_3 - m}{\beta}\right)  
    +\left. \arctan\left( \frac{r_2 - m}{\beta}\right)
  \right.\nonumber  \\
  & \; - \left.   \arctan\left( \frac{r_1 - m}{\beta}\right)  \right)\nonumber\\
  &- 2m\ln\left| \frac{(r_4^2 -2mr_4 +a^2)(r_2^2 -2mr_2 +a^2)}{(r_3^2 -2mr_3 + a^2)(r_1^2 -2mr_1 + a^2)} \right|.
\end{align}


\subsection{Transport phase in the Kerr spacetime}
\label{Transport phase in Kerr-spacetime}

We now obtain an expression for the phase $\vartheta^{\tau}(ct,r,\varphi)$ acquired by a photon as it follows a null geodesic in the Kerr spacetime. Within the WKB approximation, we have the eikonal equation $k^{\mu}\partial_{\mu}\vartheta ^{\tau} = 0$, which leads to the differential equation
\begin{equation}\label{eikonalequation_expand}
     k^{1}\frac{\partial \vartheta^{\tau}}{\partial r}  = k^{3}\mathcal{R} - k^{0}\mathcal{T}, 
\end{equation}
where $\mathcal{R}$ and $\mathcal{T}$ are conserved quantities associated with the azimutal angular momentum and the total energy, respectively (for details, see Appendix \ref{Appendix:kerr_geodesic}). 
After integration, we obtain for the transport phase
\begin{equation}
    \vartheta^{\tau} = \mathcal{T} ct + S(r, \varphi)
    \label{theta_tau_phase}
\end{equation}
with
\begin{equation}
S(r, \varphi)=-\mathcal{R}\varphi  + \int dr\frac{1}{k^1}\left(k^{3}\mathcal{R} -k^{0}\mathcal{T}\right).
\end{equation}
It is possible to show using Eq.~\eqref{eq:null_tetrad_relation} that
\begin{equation}
    k_{\mu}\xi^{\mu} = \frac{\partial \vartheta^{\tau}}{\partial x^{0}} = k^{\hat{0}}\sqrt{g_{00}}\equiv \mathcal{T},
\end{equation}
where $k^{\hat{0}}$ is the component of the null vector $k^{\mu} = dx^{\mu}/d\lambda$ on the basis of the tetrad. The quantity $c\tau=\sqrt{g_{00}}ct = \mathcal{T}ct/k^{\hat{0}}$ corresponds to the proper time measured by a static observer at a point in spacetime. The function $S(r,\varphi)$ in \eqref{theta_tau_phase}  depends only on the spatial coordinates, so in the context of interference, the phase difference will be given by the arrival time difference of the photon, that is,
\begin{equation}\label{phase-shift-kerr}
    \Delta\vartheta^{\tau} = k^{\hat{0}}c\Delta \tau =k^{\hat{0}}c\sqrt{g_{00}}\Delta t,
\end{equation}
where all functions are evaluated at the point where the photon interferes.

\subsection{Phase difference in a geodesic interferometer at the weak field and slow rotation limits}
\label{Arrival time difference in a geodesic interferometer at the weak field and slow rotation limits}

\begin{figure}[t!]
        \centering
        \includegraphics[width=8cm]{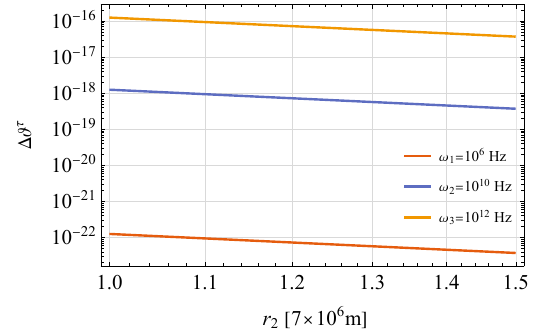}
        \caption{Phase difference $\Delta \vartheta^{\tau}$ from Eq.~\eqref{eq:theta_final_expanded_main} as a function of $r_2$ for $\omega_1 = 10^{6}\,\rm{Hz}$ (red line), $\omega_2 = 10^{10}\,\rm{Hz}$ (blue line) and $\omega_3 = 10^{12}\,\rm{Hz}$ (yellow line), with $l=0.3$ m and $h=0.7$ m.}
        \label{fig:gravitational-phase-shift-rigid-translation}
\end{figure}

In order to study a geodesic interference experiment near Earth, we use the geodesic solutions for $a>m$.
As we can observe from Fig.~\ref{fig:Geodesic_Interferometer}, the shape of the interferometer is non-trivial, as are the expressions for the flight times and the Wigner phase in the weak-field and slow-rotation approximations. To simplify the analysis of these expressions, we propose a parameterization as a function of the coordinate $r_2$ where the source is located, that is,
\begin{align}\label{parameterization-geodesic-interferometer}
    r_3 &=  r_2+ h , \; \quad  h>0, \\
    r_1 & = r_2 - l, \;\quad \; l>0,\\
    r_4(r_1,r_2,r_3) &= r_4(r_2 -l ,  r_2,  h +  r_2),
\end{align} 
and expand the arrival time difference Eq.~\eqref{arrivals_times_difference} in powers of $\left(m/r_2\right)$ and $\left(a/r_2\right)$ up to the fourth order, as the induced gravitational phase shift in Eq.~\eqref{phase-shift-kerr} depends on the arrival time difference in Eq.~\eqref{arrivals_times_difference}. We also define the phase $\phi_E$ that cancels out all contributions that are not functions of $m$ and $a$. Thus, the phase difference $\Delta\vartheta^{\tau}$ associated with the arrival time difference $\Delta\tau$ becomes
\begin{eqnarray}
 \Delta\vartheta^{\tau} &=& \frac{\omega}{c} h l \left( \frac{4 a^{2}}{r_2^3}+\frac{4a^{2}m(3 - \cos^2\theta)}{ r_2^4}\right.
 \nonumber\\
 &&\left.  +\frac{6 a^2 m^{2} (5-2 \cos ^2\theta  )}{r_2^5} -\frac{4 a^4  }{ r_2^5}\right).
 \label{eq:theta_final_expanded_main}
\end{eqnarray}
Figure\thinspace\ref{fig:gravitational-phase-shift-rigid-translation} shows $\Delta \vartheta^{\tau}$ as a function of the initial position of the photon source $r_2\in[7 \times 10^6\,{\rm m},7.5\times 10^6\,{\rm m}] $ for a geodesic interferometer with $l=0.3$ m and $h=0.7$ m for $\omega_1 = 10^{6} \, \rm{Hz} $, $\omega_2 = 10^{10} \, \rm{Hz}$ and $\omega_3 = 10^{12} \, \rm{Hz}$. Clearly, the phase difference generated by the arrival time difference exhibits very small values in the interval $[10^{-22},10^{-16}]$. The choice of values allows comparison with the study of the Kerr frame dragging effect using the Hong-Ou-Mandel dip \cite{10.1116/5.0073436}.

\subsection{Wigner phase difference in a geodesic interferometer at the weak field and slow rotation limits}
\label{Wigner Phase difference in a geodesic interferometer at the weak field and slow rotation limits}

The Wigner phase $\Delta \vartheta$ in the geodesic interferometer can be obtained by evaluating Eq.~\eqref{wigner_phase_between_two_observers} along each path of the interferometer. The Wigner phase $\vartheta_{\gamma_1,\gamma_2}$ is obtained by adding the Wigner phases $\vartheta_{\gamma_1}$ and $\vartheta_{\gamma_2}$ in paths $\gamma_1$ and $\gamma_2$, respectively. Analogously for $\vartheta_{\gamma'_1,\gamma'_2}$. Thus, the Wigner phase difference is given by
\begin{equation}
    \Delta \vartheta = \vartheta_{\gamma_1,\gamma_2} - \vartheta_{\gamma'_1,\gamma'_2}.
\end{equation}
At the weak field and slow rotation limits, the Wigner phase difference can be approximated as
\begin{eqnarray}
\Delta \vartheta &=&  \frac{ahlm}{r_2^4}\left[ \frac{\cos\theta}{r_2^2}(10a^2 + 6a^2\cos2\theta )\right.
\nonumber\\
& & \left. - \; 4\cos\theta\left(1+\frac{2m}{r_2}\right) \right].
\label{eq:Wigner-phase-interferometer}
\end{eqnarray}

\begin{figure}[t!]
        \centering
        \includegraphics[width=8cm]{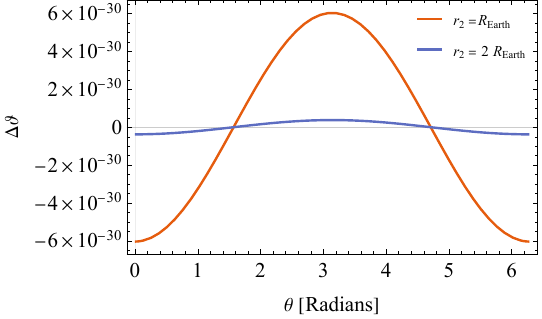}
        \caption{Wigner phase rotation $\Delta\varphi$ from Eq.~\eqref{eq:Wigner-phase-interferometer} as a function of the angular coordinate $\theta$. Each line represents the distance $r_2$ of the source of photons from the center of the Earth for $\Delta r=r_3-r_1=1$ m. The red continuous line corresponds to $r_2 = R_{\rm{Earth}}$ and the blue continuous line to $r_2 = 2 \, R_{\rm{Earth}}$.}
        \label{fig:Wigner-phase-small-interferometer}
    \end{figure}

\begin{figure}[t!]
        \centering
        \includegraphics[width=8cm]{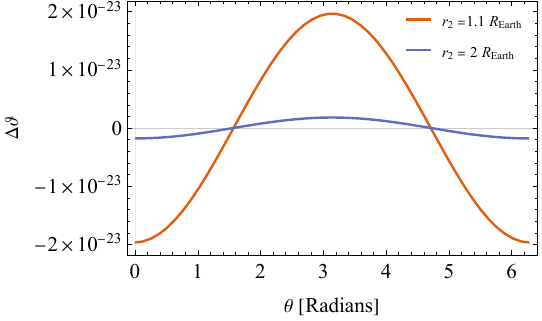}
        \caption{Wigner phase rotation $\Delta\varphi$ from Eq.~\eqref{eq:Wigner-phase-interferometer} as a function of the angular coordinate $\theta$. Each line represents the distance $r_2$ of the source of photons from the center of the Earth for $\Delta r=r_3-r_1 \approx 2 \times 10^3$ m. The red continuous line corresponds to $r_2 = 1.1 \,R_{\rm{Earth}}$ and the blue continuous line to $r_2 = 2 \, R_{\rm{Earth}}$.}
        \label{fig:Wigner-phase-big-interferometer}
    \end{figure}

Figure~\ref{fig:Wigner-phase-small-interferometer} shows $\Delta\vartheta$ as a function of the angular coordinate $\theta$ for the case of Earth with $R_{\rm Earth}\approx7\times10^6\,{\rm m}$ and $\Delta r := r_{3}-r_{1} \approx 1\,{\rm m}$ with $r_2=R_{\rm Earth}$ (red line) and $r_2=2R_{\rm Earth}$ (blue line). As is apparent from this figure, the Wigner phase $\Delta\vartheta$ reaches values of the order of $10^{-30}$, which is several orders of magnitude smaller than the phase shift $\Delta\vartheta^\tau$ generated by the arrival time difference. The value of $\Delta\vartheta$ can be increased by considering higher values of $\Delta r$. This is shown in Fig.~\ref{fig:Wigner-phase-big-interferometer}, where $\Delta r \approx 2 \times 10^3\,{\rm m}$, leading to Wigner phase values on the order of $10^{-23}$.

\subsection{Detection probability and visibility in a geodesic interferometer}
\label{Detection probability and visibility in a geodesic interferometer}

We consider now the propagation of a single photon in the geodesic interferometer. The single photon is described by the state $|\psi\rangle$ of Eq.~\eqref{initial_photon_state}, which considers frequency dispersion and helicity, before entering the interferometer. After propagation, the photon is measured at both output ports of the interferometer at $r_4$. We assume that the measurement devices  do not provide information about the four-momentum $\hat{k}$ or helicity $s$. Thus, we model the measurement through the operators $\Pi_+ = \int d\hat{\mathbf{k}} f^{\dagger}_{\hat{\mathbf{k}}} \vert 0 \rangle \langle 0 \vert f_{\hat{\mathbf{k}}}$ and $\Pi_- = \int d\hat{\mathbf{k}} g^{\dagger}_{\hat{\mathbf{k}}} \vert 0\rangle \langle 0\vert g_{\hat{\mathbf{k}}}$, which lead to the detection probability (for details, see Appendix \ref{Appendix:Probability}) 
\begin{align}
    P_{\pm} = \frac{1}{2}\left[1 \pm \int d\hat{\mathbf{k}} \vert f(\hat{\mathbf{k}}) \vert^2 \cos \left( \Delta\vartheta \right)\cos\left( \Delta\vartheta^{\tau} +\phi_{E}\right) \right].
    \label{ProbMZ1}
\end{align}
As expected, this expression contains an armonic dependence on the phase difference $\Delta\vartheta^\tau$, which is associated with the difference of arrival time, and the accumulated phase $\phi_{E}$, which depends on the action of the mirrors and controllable local phases. However, there is another armonic dependence on the Wigner phase difference $\Delta\vartheta$. Probability $P_\pm$ exhibits a similar dependence as in the case of massive particles with inner degrees of freedom \cite{Zych2011,Basso2021a}.

If we assume that the function  $f(\hat{\mathbf{k}})$ is a Gaussian distribution centered in $\omega$ with spectral width $\sigma$ then the probability of detection $P_\pm$ Eq.~\eqref{ProbMZ1} becomes
\begin{equation}
    P_{\pm}=\frac{1}{2}[1 \pm V \cos{\left( \Delta \vartheta^{\tau}+\phi_E\right)}],
    \label{ProbMZ2}
\end{equation}
where the interferometric visibility $V=V_{\rm \vartheta^{\tau}}V_{\rm \vartheta}$ is given by the functions
\begin{equation}
    V_{\vartheta^{\tau}} = e^{-(\frac{\sigma}{\omega}\Delta\vartheta^{\tau})^2}
    \label{eq:V_tau}
\end{equation}
and
\begin{equation}
    V_{\vartheta} = \cos\left(\Delta \vartheta \right).
    \label{eq:V_theta}
\end{equation}
The component $V_{\vartheta^{\tau}}$ exponentially reduces the detection probability as $\Delta\vartheta^{\tau}$ increases, for fixed $\omega$ and $\sigma$. The other component, $V_{\vartheta}$ given by Eq.~\eqref{eq:Wigner-phase-interferometer}, arises from the Wigner phase and shows the coupling of the metric elements, such as $a$ and $m$ in the Kerr metric, to the photon helicity. If we neglect this gravitational coupling, i.e., $\Delta\vartheta = 0$ and $V_{\vartheta} = 1$), the probability simplifies to $P_{\pm} = (1 \pm V_{\vartheta^{\tau}} \cos(\Delta \vartheta^{\tau}))/2$ in agreement with previously obtained results \cite{Zych_2012}.

\subsection{Estimation of the specific angular momentum}
\label{Estimation of the specific angular momentum}
\begin{figure}[t!]
    \centering 
    \includegraphics[width=0.45\textwidth]{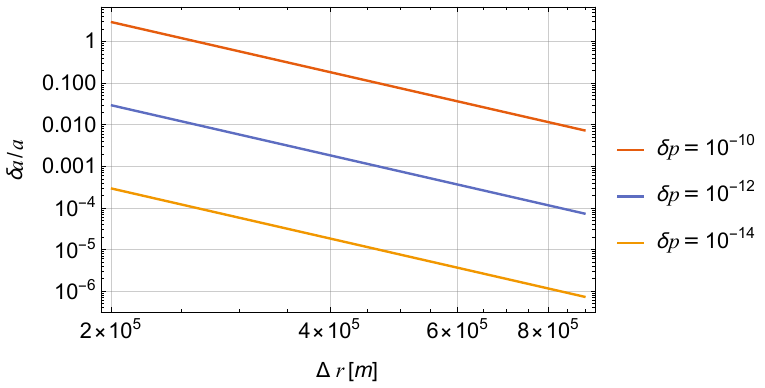} 
   \caption{Relative error $\delta a/a$, with $a$ estimated from Eq.~\eqref{a-estimation-Gamma-terms}, as a function of the distance $\Delta r=r_3-r_1$ for $\delta p = 10^{-10}$ (red continuous line), $\delta p = 10^{-12}$ (blue continuous line) and $\delta p = 10^{-14}$ (yellow continuous line) with $\omega = 10^{15}$ Hz and $\sigma = 10^{13}$ Hz.} 
    \label{fig:Rel_Error_a_Small_Interferometer} 
\end{figure}

The detection probability can be used to estimate the specific angular momentum $a$ that defines the Kerr spacetime metric. 
We assume that the phase difference $\Delta \vartheta^{\tau}$ and the Wigner phase difference $\Delta \vartheta$ are sufficiently small. Furthermore, we assume that the term $(\sigma/\omega) \Delta\vartheta^{\tau}$ is also considered small. Under these assumptions, a second-order Taylor expansion of probability $P_{+}$ Eq.~\eqref{ProbMZ2} yields
\begin{equation} 
    P_{+} = \frac{1}{4} [(\Delta\vartheta)^2 + \left(1+\frac{\sigma^2}{\omega^2}\right) (\Delta\vartheta^{\tau})^2 ]+ \mathcal{O}(3),
\label{eq:prob_approx}
\end{equation}
where $\mathcal{O}(3)$ denotes all terms of cubic order or higher. Inserting in $P_+$ the expressions for $\Delta \vartheta^{\tau}$ and $\Delta \vartheta$ as functions of $a$, we obtain 
\begin{equation}
    P_{+} = 1-a^4 \left(\frac{8 h^2 l^2 \sigma ^2}{c^4 r_2^6}+\frac{4 h^2 l^2 \omega ^2}{c^4 r_2^6}\right)-a^2\frac{ \left(4 m^2 h^2 l^2 \cos ^2\theta \right)}{r_2^8}.
    \label{prob_Aprox_To_Solve}
\end{equation}
This can be solved algebraically for the specific angular momentum $a$ yielding 
\begin{widetext}
\begin{equation}
    a = \pm \frac{1}{\sqrt{2}}\sqrt{ \frac{c^2  \sqrt{c^4 h^2 l^2 m^4 \cos ^4(\theta )-(P_{+}-1) r_2^{10} \left(2 \sigma ^2+\omega ^2\right)}}{h l r_2^2 \left(2 \sigma ^2+\omega ^2\right)} - \frac{c^4 m^2 \cos ^2\theta }{ r_2^2 \left(2 \sigma ^2+\omega ^2\right)} }.
    \label{a-estimation-Gamma-terms}
\end{equation}
\end{widetext}
This expression can be used to estimate the value of the specific angular momentum $a$ from the value of the probability $P_+$. To assess the uncertainty of the estimate, we calculate the relative error $\delta a/a$ using standard error propagation (for further details, see Appendix \ref{Appendix:Error analysis}). This is shown in Fig.~\ref{fig:Rel_Error_a_Small_Interferometer} as a function of the size $\Delta r := r_3 - r_1$ of the interferometer for three different values of the error in the value of the probability $\delta p = 10^{-14}$ (yellow line), $\delta p = 10^{-12}$ (blue line) and $\delta p = 10^{-10}$ (red line) with $r_2 = 1.05 \times 10^7$ m. The plot shows that the relative error decreases as $\Delta r$ increases, reaching an order of magnitude of $10^{-6}$ for a probability uncertainty of $\delta p = 10^{-14}$. 

\section{Conclusion}

We have studied the rotation of the polarization due to the gravitational field in the case of Kerr spacetime. Our approach was based on the definition of a geodesic interferometer, that is, a Mach-Zehnder interferometer whose arms are defined by the propagation of light along null geodesics. In particular, we have defined the interferometer by setting the azimuthal coordinate as constant, which allows us to analytically calculate null geodesics, where the radial coordinate plays the role of an affine parameter. We consider the case of a single photon propagating within the interferometer, whose quantum state includes frequency dispersion, and we obtain the detection probability at the output ports of the interferometer. This is an armonic function of the phase difference $\Delta\vartheta^\tau$ that arises due to the arrival time difference between the arms of the interferometer, which in turn is controlled by the temporal gravitational delay. The detection probability is also characterized by an interferometric visibility that decreases quadratically with $\Delta\vartheta^\tau$ and is modulated by an armonic function that depends on the Wigner phase $\Delta\vartheta$. Thus, the detection probability is a signature of two purely relativistic effects acting on a quantum system. This result is valid for the limits of slow rotation and weak gravitational field. The exponential decrease in visibility due to the phase shift $\Delta\vartheta^\tau$ is consistent with the results in \cite{Zych_2012}, while the drop due to rotation of the polarization parallels the case for massive particles \cite{Zych2011}. For a geodesic interferometer on Earth's surface and with a radial coordinate interval between mirrors of the order of 2 km, $\Delta\vartheta^\tau$ and $\Delta\vartheta$ have values close to $10^{-16}$ and $10^{-23}$, respectively. Finally, we have also studied the estimation of the specific angular momentum from the detection probability and obtained an expression for the uncertainty of the estimate as a function of the uncertainty in the detection probability. For a geodesic interferometer near Earth, with mirrors separated by a radial coordinate distance of about 800 km and an error of $10^{-14}$ for the detection probability, the resulting relative uncertainty in determining the specific angular momentum is on the order of $10^{-6}$.

\section*{Acknowledgements}
AD and FJL were supported by the National Agency of Research and Development (ANID) -- Millennium Science Initiative Program -- ICN17$_-$012. MRT acknowledges the support of ANID through Fondecyt grant No. 3230407.

\newpage
\appendix
\begin{widetext}

\section{Geodesics in the Kerr spacetime} 
\label{Appendix:kerr_geodesic}
In Schwarzschild spacetime it is possible to define radial null geodesics \cite{MTW1973}. However, Kerr spacetime does not allow their existence, since it lacks spherical symmetry. Nevertheless, it is possible to find analytical solutions in the particular case where the trajectories are restricted to a cone defined by $\theta = cte$. 

Geodesic equations can be obtained from the Lagrangian
\begin{equation}
    \widetilde{\mathcal{L}} = \frac{1}{2} g_{\mu \nu} \dot{x}^{\mu} \dot{x}^{\nu},
\end{equation}
where $\dot{x}^{\mu} = dx^{\mu}/d\lambda$, with $\lambda$ an affin parameter. By expanding the above equation using the Kerr metric \eqref{eq:KerrComponents}, we obtain the expression
\begin{equation}
    2\widetilde{\mathcal{L}} = \frac{\Delta}{\rho^{2}}\left( c \dot{t} - a \sin^{2}{\theta} \dot{\phi}\right)^{2} - \frac{\sin^{2}\theta}{\rho}\left[\left(r^{2} + a^{2}\right)\dot{\phi} - ac\dot{t}\right]^{2}
    -\frac{\rho ^{2}}{\Delta}\dot{r}^{2}.
\end{equation}
The Euler-Lagrange equations, given by
\begin{equation}
    \frac{\partial \widetilde{\mathcal{L}} }{\partial x^{\mu}} - \frac{d}{d \lambda}\left( \frac{ \partial \widetilde{\mathcal{L}}}{\partial \dot{x}^{\mu}}\right) = 0,
\end{equation}
for $x^{0} = c t$ lead to
\begin{equation} \label{cantidadconservada_l}
    \frac{d}{d\lambda}\left(  \frac{ \partial \widetilde{\mathcal{L}}}{\partial c\dot{t}}\right) = 0 \, \, \, \, \, \, \, \Rightarrow \, \frac{ \partial \widetilde{\mathcal{L}}}{\partial c\dot{t}} = cte := \mathcal{T},
\end{equation}
where $\mathcal{T}$ is an integration constant associated with the energy. Expanding Eq.~\eqref{cantidadconservada_l} we have
\begin{equation}
    \frac{\Delta}{\rho^{2}}\left( c \dot{t} - a \sin^{2}{\theta} \dot{\phi}\right) + a\frac{\sin^{2}\theta}{\rho^{2}}\left[\left(r^{2} + a^{2}\right)\dot{\phi} - ac\dot{t}\right] = \mathcal{T}.
\end{equation}
Now, for $x^{3}=\varphi$, the Euler-Lagrange equations become 
\begin{equation} \label{cantidadconservada_n}
    \frac{d}{d\lambda}\left(  \frac{ \partial \widetilde{\mathcal{L}}}{\partial \dot{\varphi}}\right) = 0 \, \, \, \, \, \, \, \Rightarrow \, \frac{ \partial \widetilde{\mathcal{L}}}{\partial \dot{\varphi}} = cte := -\mathcal{R}, 
\end{equation}
where $\mathcal{R}$ is an integration constant associated with the angular momentum. Expanding Eq.~\eqref{cantidadconservada_n} we have
\begin{equation}
    a \sin^{2}{\theta} \,\frac{\Delta}{\rho^{2}}\left( c \dot{t} - a \sin^{2}{\theta} \dot{\phi}\right) + \left(r^{2} + a^{2}\right)\frac{\sin^{2}\theta}{\rho^{2}}\left[\left(r^{2} + a^{2}\right)\dot{\phi} - ac\dot{t}\right] = \mathcal{R}.
\end{equation}
Analogously, for $x^{2}= \theta$ we obtain the equation
\begin{equation}
    a^{2} \frac{\Delta}{\rho^{4}}\left( c \dot{t} - a \sin^{2}{\theta} \dot{\varphi}\right)^{2} 
    -2a \dot{\varphi} \frac{\Delta}{\rho^{2}}\left( c \dot{t} - a \sin^{2}{\theta} \dot{\varphi}\right)
     + \frac{\left(r^{2} + a^{2}\right)}{\rho^{4}}\left[\left(r^{2} + a^{2}\right)\dot{\varphi} - ac\dot{t}\right]^{2} + \frac{a^{2}\dot{r}^{2}}{\Delta} = 0.
\end{equation}

It can be seen that there are only three unknowns $(\dot{t},\dot{r},\dot{\phi})$ and four equations. Thus, there must exist a relation between the constants $\mathcal{T}$ and $\mathcal{R}$.  Multiplying by $\left(r^{2} + a^{2}\right)$ the equations for $\mathcal{T}$ and $\mathcal{R}$, we obtain the equation
\begin{equation}
    \left(r^{2} + a^{2}\right)\frac{\Delta}{\rho^{2}}\left( c \dot{t} - a \sin^{2}{\theta} \dot{\phi}\right) + a\left(r^{2} + a^{2}\right)\frac{\sin^{2}\theta}{\rho^{2}}\left[\left(r^{2} + a^{2}\right)\dot{\phi} - ac\dot{t}\right] 
     = \left(r^{2} + a^{2}\right)\mathcal{T}.
\end{equation}
Noting that the second term of this equation also appears in the expression for $\mathcal{R}$, we obtain
\begin{equation}
    \left(r^{2} + a^{2}\right)\frac{\Delta}{\rho^{2}}\left( c \dot{t} - a \sin^{2}{\theta} \dot{\phi}\right) + a\left[ \mathcal{R} - a \sin^{2}{\theta} \,\frac{\Delta}{\rho^{2}}\left( c \dot{t} - a \sin^{2}{\theta} \dot{\phi}\right) \right] =\left(r^{2} + a^{2}\right)\mathcal{T},
\end{equation}
which after rearrenging terms becomes
\begin{equation}
    \frac{\Delta}{\rho^{2}}\left( c \dot{t} - a \sin^{2}{\theta} \dot{\phi}\right)\left[ \left(r^{2} + a^{2}\right) - a^{2} \sin^{2}{\theta}  \right] =\left(r^{2} + a^{2}\right)\mathcal{T} -a \mathcal{R}.
\end{equation}
Recalling the definition $\rho^{2} = r^{2} + a^{2}\cos^{2}{\theta}$, the previous expression can be cast in compact form
\begin{equation}\label{I}
    \Delta\left( c \dot{t} - a \sin^{2}{\theta} \dot{\phi}\right) =\left(r^{2} + a^{2}\right)\mathcal{T} -a \mathcal{R}.
\end{equation}
A similar procedure can be applied to the equation for $\mathcal{T}$. Multiplying it by $ a\sin^{2}{\theta}$  we get
\begin{equation}
    \frac{\Delta a\sin^{2}{\theta}}{\rho^{2}}\left( c \dot{t} - a \sin^{2}{\theta} \dot{\phi}\right) + a^{2}\frac{\sin^{4}\theta}{\rho^{2}}\left[\left(r^{2} + a^{2}\right)\dot{\phi} - ac\dot{t}\right] = a\sin^{2}{\theta}\mathcal{T},
\end{equation}
which after substituting the equation for $\mathcal{R}$ leads to
\begin{eqnarray}\label{II}
    \sin^{2}\theta\left[\left(r^{2} + a^{2}\right)\dot{\phi} - ac\dot{t}\right] = \mathcal{R}-a\sin^{2}{\theta}\mathcal{T}.
\end{eqnarray}
Adding the results of multiplying Eq.~\eqref{I} and \eqref{II} by $a\sin^{2}\theta$ and $\Delta$, respectively, we obtain an equation that relates the conserved quantities $\mathcal{T}$ and $\mathcal{R}$, that is,
 \begin{equation} \label{conserved_relation}
     \left(\mathcal{R} + a\, \mathcal{T}\, \sin^{2}(\theta)\right)\left(\mathcal{R} - a\, \mathcal{T}\, \sin^{2}(\theta)\right)=0.
 \end{equation}
Inserting the solution $\left(\mathcal{R} - a\, \mathcal{T}\, \sin^{2}(\theta)\right)=0$ in Eq.~\eqref{II}, we obtain
\begin{equation}\label{t_punto}
    c\dot{t} = \dot{\varphi}\frac{\left(r^{2} +a^{2}\right)}{a}.
\end{equation}
Substituting Eq.~\eqref{t_punto} into the equation for $\mathcal{R}$, we get 
\begin{equation}
    \dot{\varphi} = \frac{\mathcal{R}}{\Delta \sin^{2}\theta}.
\end{equation}
Using the two previous results, it is possible to find an equation for $c\dot{t}$ as a function of $r$, that is,
\begin{equation}
    c\dot{t} = \frac{\left(r^{2} +a^{2}\right)}{a}\frac{\mathcal{R}}{\Delta \sin^{2}\theta}.
\end{equation}
By substituting the three previous results into the equation $ds^{2}=0$ for null geodesics, we obtain the differential equation
\begin{equation}
    \dot{r}^{2} = \frac{\mathcal{R}^{2}}{a^{2}}\frac{1}{\sin^{4}\theta}
    \Rightarrow \dot{r} = \pm \frac{\mathcal{R}}{a} \frac{1}{\sin^{2}\theta},
\end{equation}
whose solutions are
\begin{equation}
    r(\lambda) = \pm \frac{\mathcal{R}}{a\sin^{2}\theta} \, \lambda + cte.
\end{equation}
This result gives us the freedom to choose $r$ as the parameter along each geodesic. In particular, by selecting the positive solution for $\dot{r}$, we obtain:
\begin{equation}\label{eq:A-dphidr}
    \frac{d\varphi}{d\lambda} = \frac{dr}{d\lambda} \frac{d\varphi}{dr} = \frac{\mathcal{R}}{a\sin^{2}\theta}\frac{d\varphi}{dr}=\frac{\mathcal{R}}{\Delta \sin^{2}\theta}
    \Rightarrow \frac{d\varphi}{dr} = \frac{a}{\Delta}.
\end{equation}
Analogously, for $dt/d\lambda$ we have the expression
\begin{equation}\label{eq:A-dctdr}
    c\frac{dt}{d\lambda} = \frac{dr}{d\lambda} c\frac{dt}{dr}=\frac{\mathcal{R}}{a\sin^{2}\theta}c\frac{dt}{dr} = \left(r^{2} +a^{2}\right)\frac{\mathcal{R}}{a^{2} \sin^{2}\theta}
    \Rightarrow c\frac{dt}{dr} =  \frac{\left(r^{2} +a^{2}\right)}{\Delta}.
\end{equation}
Integrating the expressions \eqref{eq:A-dphidr} and \eqref{eq:A-dctdr} considering $a>m$ we obtain the outgoing and ingoing solutions
\begin{equation}
    \varphi(r)_{\pm} = \pm\frac{a}{\sqrt{a^2 -m^2}}\left(\arctan\left( \frac{r-m}{\sqrt{a^2 - m^2}}\right) - \arctan\left(\frac{r_0 - m}{\sqrt{a^2 - m^2}} \right)\right) + \varphi_0,
    \label{phi_Geodesic_new}
\end{equation}
\begin{equation}
    ct(r)_{\pm} = \pm\Delta r \pm \frac{2m^2}{\sqrt{a^2 - m^2}}\left(\arctan\left( \frac{r-m}{\sqrt{a^2 - m^2}}\right) - \arctan\left(\frac{r_0 - m}{\sqrt{a^2 - m^2}} \right) \right) \pm m \ln\left| \frac{r^2 -2mr +a^2}{r_0^2 -2mr_0 + a^2}\right| + ct_0.
    \label{ct_Geodesic_new}
\end{equation}

The outgoing solutions for $m>a$ are given by:
\begin{eqnarray} \label{null_geodesic_phi_out}
\varphi_{\pm}(r)=\pm\frac{a}{2\sqrt{m^{2} - a^{2}}}\ln\left| \frac{(r-r_{+})}{(r - r_{-})}\frac{(r_0-r_{-})}{(r_0-r_{+})}\right| + \varphi_0
\end{eqnarray}
and 
\begin{eqnarray}
c t_{\pm}(r)= \pm\Delta r \pm m_+\ln\left| \frac{(r-r_{+})}{(r_0 - r_{+})}\right|    \pm m_-\ln\left| \frac{(r-r_{-})}{(r_0 - r_{-})}\right| + ct_0,
\end{eqnarray}
where
\begin{equation}
    m_{\pm} = \left( m \pm \frac{m^{2}}{\sqrt{m^{2}-a^{2}}}\right) \; \; \; {\rm and }\; \; \; r_{\pm} = m \pm \sqrt{ m^{2} - a^{2}}.
\end{equation}


Using Eq. \eqref{null_geodesic_phi_out} for the geodesic interferometer, we find that the interference coordinate $r_4$ coincides with the expression given in Eq. \eqref{r4} of the main text.
Following the section \ref{Geodesic interferometer}, the arrival proper time difference $c\Delta\tau=c(\tau'-\tau)$ between both paths of the interferometer in the case $m>a$ is given by
\begin{align}\label{arrivals_times_difference_m>a}
  c\Delta \tau =2\left(r_3 + r_1 -r_4-r_2\right)
  +2m_+\ln\left| \frac{(r_3-r_{+})(r_1-r_{+})}{(r_4 - r_{+})(r_2 - r_{+})}\right| 
   +2m_-\ln\left| \frac{(r_3-r_{-})(r_1-r_{-})}{(r_4 - r_{-})(r_2 - r_{-})}\right| .
\end{align}
with $\Pi_4$ as in the main text.

\section{Detection probability in the geodesic interferometer}
\label{Appendix:Probability}

A single photon before entering the inteferometer is described by the state
\begin{equation}
    \ket{\psi} = \frac{1}{\sqrt{2}} \sum_{s = \pm 1} \int d\hat{\mathbf{k}} f(\hat{\mathbf{k}}) \ket{\hat{\mathbf{k}},s}.
\end{equation}
After the action of the first beam splitter, the state becomes
\begin{align}
\vert \Psi \rangle = \frac{1}{2}\sum_{s= \pm 1} \int d\hat{\mathbf{k}} f(\hat{\mathbf{k}})\left[  e^{i \vartheta^{\tau}_{\gamma_1}} e^{i s\vartheta_{\gamma_1}}\hat{a}^{\dagger}_{\hat{\mathbf{k}}} \vert 0_{\hat{\mathbf{k}}}, s  \rangle +  e^{-i\phi_{BS_{1}}} e^{i s\vartheta_{\gamma_1'}} e^{i \vartheta^{\tau}_{\gamma_1'}} \hat{b}^{\dagger}_{\hat{\mathbf{k}}} \vert 0_{\hat{\mathbf{k}}}, s  \rangle \right], \nonumber
\end{align}
where $\vartheta^{\tau}_{\gamma_j}$ and $\vartheta_{\gamma_j}$ are the phases that originate in the gravitational time delay and  the Wigner phase along the path $j$, respectively, and $\hat{a}^{\dagger}_{\hat{\mathbf{k}}}$ and $\hat{b}^{\dagger}_{\hat{\mathbf{k}}}$ are bosonic creation operators in propagation modes $\gamma_1$ and $\gamma'_1$, respectively. When the photon is reflected in mirrors at $r_1$ and $r_3$, the state becomes
\begin{equation}
    \vert \Psi \rangle = \frac{1}{2}\sum_{s= \pm 1} \int d\hat{\mathbf{k}} f(\hat{\mathbf{k}}) \left[  e^{i (\vartheta^{\tau}_{\gamma_1}  + \vartheta^{\tau}_{\gamma_2})} e^{i s(\vartheta_{\gamma_1} + \vartheta_{\gamma_2})}\hat{a}^{\dagger}_{\hat{\mathbf{k}}} \vert 0_{\hat{\mathbf{k}}}, s  \rangle 
     +  e^{-i\phi_{E}} e^{i s\left(\vartheta_{\gamma_1'} + \vartheta_{\gamma_2'}\right)} e^{i \left(\vartheta^{\tau}_{\gamma_1'} + \vartheta^{\tau}_{\gamma_2'}\right)} \hat{b}^{\dagger}_{\hat{\mathbf{k}}} \vert 0_{\hat{\mathbf{k}}}, s  \rangle \right]. 
\end{equation}
After the last beam splitter, the state of the single photon becomes
\begin{equation}
    \vert \Psi_{\rm out} \rangle =  \frac{1}{2\sqrt{2}}\sum_{s= \pm 1} \int d\hat{\mathbf{k}} f(\hat{\mathbf{k}}) \left[  e^{i \Theta_{\hat{\mathbf{k}}}} e^{i s\Omega_{\hat{\mathbf{k}}}}\left(\hat{f}^{\dagger}_{\hat{\mathbf{k}}} + \hat{g}^{\dagger}_{\hat{\mathbf{k}}}\right) \vert 0_{\hat{\mathbf{k}}}, s  \rangle 
     +  e^{-i\phi_{E}} e^{i \Theta'_{\hat{\mathbf{k}}}} e^{is \Omega'_{\hat{\mathbf{k}}}} \left(\hat{f}^{\dagger}_{\hat{\mathbf{k}}} - \hat{g}^{\dagger}_{\hat{\mathbf{k}}}\right) \vert 0_{\hat{\mathbf{k}}}, s  \rangle \right],
\end{equation}
where for simplicity, we have defined the quantities,
\begin{align}
   \Theta_{\hat{\mathbf{k}}} \equiv&  \,\Theta({\hat{\mathbf{k}}})= \, \vartheta^{\tau}_{\gamma_1}({\hat{\mathbf{k}}})  + \vartheta^{\tau}_{\gamma_2}({\hat{\mathbf{k}}}),\\
   \Omega_{\hat{\mathbf{k}}} \equiv&  \,\Omega({\hat{\mathbf{k}}})= \,\vartheta_{\gamma_1}({\hat{\mathbf{k}}}) + \vartheta_{\gamma_2}({\hat{\mathbf{k}}}),
\end{align}
and $\hat{f}^{\dagger}_{\hat{\mathbf{k}}}$ and $\hat{g}^{\dagger}_{\hat{\mathbf{k}}}$ are bosonic creation operators at the output ports of the beam splitter.

We consider detectors that respond flat, that is, they do not provide information about the vector $\hat{\mathbf{k}}$. Therefore, in this case, we consider projectors $\Pi_+ = \int d\hat{\mathbf{k}} f^{\dagger}_{\hat{\mathbf{k}}} \vert 0_{\hat{\mathbf{k}}} \rangle \langle 0_{\hat{\mathbf{k}}} \vert f_{\hat{\mathbf{k}}}$ and $\Pi_- = \int d\hat{\mathbf{k}} g^{\dagger}_{\hat{\mathbf{k}}'} \vert 0_{\hat{\mathbf{k}}'} \rangle \langle 0_{\hat{\mathbf{k}}'} \vert g_{\hat{\mathbf{k}}'}$. The probability of detection of a photon in one of the output ports is given by
\begin{align}\label{ProbMZ}
    P_{+} =& \langle \Psi_{\rm out} \vert \Pi_+  \vert \Psi_{\rm out} \rangle, \nonumber\\
          =&  \frac{1}{8}\sum_{s= \pm 1} \sum_{s'= \pm 1} \int \int  d\hat{\mathbf{k}} f^{*}(\hat{\mathbf{k}})  d\hat{\mathbf{k}'} f(\hat{\mathbf{k}'})\left[  \langle s' , 0_{\hat{\mathbf{k}'}}\vert\left(\hat{f}_{\hat{\mathbf{k}'}} + \hat{g}_{\hat{\mathbf{k}'}}\right) e^{-i \Theta^{*}_{\hat{\mathbf{k}'}}} e^{-i s\Omega^{*}_{\hat{\mathbf{k}'}}}   \right.\nonumber \\
    & + \left.  \langle s' , 0_{\hat{\mathbf{k}'}}\vert \left(\hat{f}_{\hat{\mathbf{k}'}} - \hat{g}_{\hat{\mathbf{k}'}}\right) e^{i\phi^{*}_{E}} e^{-i \Theta'^{*}_{\hat{\mathbf{k}'}}} e^{-is \Omega'^{*}_{\hat{\mathbf{k}'}}} \right] \int d\bar{\hat{\mathbf{k}}} f^{\dagger}_{\bar{\hat{\mathbf{k}}}} \vert 0_{\bar{\hat{\mathbf{k}}}} \rangle \langle 0_{\bar{\hat{\mathbf{k}}}} \vert f_{\bar{\hat{\mathbf{k}}}} \nonumber \left[e^{i \Theta_{\hat{\mathbf{k}}}} e^{i s\Omega_{\hat{\mathbf{k}}}}\left(\hat{f}^{\dagger}_{\hat{\mathbf{k}}} + \hat{g}^{\dagger}_{\hat{\mathbf{k}}}\right) \vert 0_{\hat{\mathbf{k}}}, s  \rangle \right.\\
    & + \left. e^{-i\phi_{E}} e^{i \Theta'_{\hat{\mathbf{k}}}} e^{is \Omega'_{\hat{\mathbf{k}}}} \left(\hat{f}^{\dagger}_{\hat{\mathbf{k}}} - \hat{g}^{\dagger}_{\hat{\mathbf{k}}}\right) \vert 0_{\hat{\mathbf{k}}}, s  \rangle \right],
\end{align}
\begin{align}
    P_+=& \frac{1}{8}\sum_{s= \pm 1} \sum_{s'= \pm 1} \int \int \int  d\hat{\mathbf{k}} d\bar{\hat{\mathbf{k}}}d\hat{\mathbf{k}'} f^{*}(\hat{\mathbf{k}}) f(\hat{\mathbf{k}'})\left[  \langle s' , 0_{\hat{\mathbf{k}'}}\vert\left(\hat{f}_{\hat{\mathbf{k}'}} + \hat{g}_{\hat{\mathbf{k}'}}\right)\hat{f}^{\dagger}_{\bar{\hat{\mathbf{k}}}} \vert 0_{\bar{\hat{\mathbf{k}}}} \rangle  e^{-i \Theta^{*}_{\hat{\mathbf{k}'}}} e^{-i s\Omega^{*}_{\hat{\mathbf{k}'}}}   \right.\nonumber \\
    & + \left.  \langle s' , 0_{\hat{\mathbf{k}'}}\vert \left(\hat{f}_{\hat{\mathbf{k}'}} - \hat{g}_{\hat{\mathbf{k}'}}\right)\hat{f}^{\dagger}_{\bar{\hat{\mathbf{k}}}} \vert 0_{\bar{\hat{\mathbf{k}}}} \rangle e^{i\phi^{*}_{E}} e^{-i \Theta'^{*}_{\hat{\mathbf{k}'}}} e^{-si \Omega'^{*}_{\hat{\mathbf{k}'}}} \right]\left[e^{i \Theta_{\hat{\mathbf{k}}}} e^{i s\Omega_{\hat{\mathbf{k}}}}\langle 0_{\bar{\hat{\mathbf{k}}}} \vert f_{\bar{\hat{\mathbf{k}}}}\left(\hat{f}^{\dagger}_{\hat{\mathbf{k}}} + \hat{g}^{\dagger}_{\hat{\mathbf{k}}}\right) \vert 0_{\hat{\mathbf{k}}}, s  \rangle \right.          \nonumber \\
    & + \left. e^{-i\phi_{E}} e^{i \Theta'_{\hat{\mathbf{k}}}} e^{i s\Omega'_{\hat{\mathbf{k}}}} \langle 0_{\bar{\hat{\mathbf{k}}}} \vert \hat{f}_{\bar{\hat{\mathbf{k}}}}\left(\hat{f}^{\dagger}_{\hat{\mathbf{k}}} - \hat{g}^{\dagger}_{\hat{\mathbf{k}}}\right) \vert 0_{\hat{\mathbf{k}}}, s  \rangle \right].
\end{align}
Since we have that
\begin{align}
\langle 0_{\hat{\mathbf{k}'}}\vert\hat{f}_{\hat{\mathbf{k}'}}\hat{f}^{\dagger}_{\hat{\mathbf{k}}}\vert 0_{\hat{\mathbf{k}}}, s  \rangle = \delta^{(3)}(\hat{\mathbf{k}'} - \hat{\mathbf{k}})\vert s\rangle = \delta_{\bar{\mathbf{k}'}, \bar{\mathbf{k}}}\vert s\rangle,
\end{align}
denoting for convenience $\delta^{(3)}(\bar{\mathbf{k}'} - \bar{\mathbf{k}}) \equiv \delta_{\hat{\mathbf{k}'}, \hat{\mathbf{k}}}$, the probability becomes
\begin{align}
    P_+=& \frac{1}{8}\sum_{s= \pm 1} \sum_{s'= \pm 1} \int \int \int  d\hat{\mathbf{k}} d\bar{\hat{\mathbf{k}}}d\hat{\mathbf{k}'} f^{*}(\hat{\mathbf{k}}) f(\hat{\mathbf{k}'})\left[  \langle s'\vert   e^{-i \Theta^{*}_{\hat{\mathbf{k}'}}} e^{-i s' \Omega^{*}_{\hat{\mathbf{k}'}}} \delta_{\hat{\mathbf{k}'}, \bar{\hat{\mathbf{k}}}}  \right.\nonumber \\
    & + \left.  \langle s' \vert e^{-i\phi^{*}_{E}} e^{-i '\Theta'^{*}_{\hat{\mathbf{k}'}}} e^{-i s\Omega'^{*}_{\hat{\mathbf{k}'}}}\delta_{\hat{\mathbf{k}'}, \bar{\hat{\mathbf{k}}}} \right]\left[\delta_{\hat{\mathbf{k}}, \bar{\hat{\mathbf{k}}}}e^{i \Theta_{\hat{\mathbf{k}}}} e^{i s\Omega_{\hat{\mathbf{k}}}}  \vert s \rangle  + \delta_{\hat{\mathbf{k}}, \bar{\hat{\mathbf{k}}}}e^{i\phi_{E}} e^{i \Theta'_{\hat{\mathbf{k}}}} e^{i s\Omega'_{\hat{\mathbf{k}}}} \vert s  \rangle \right],
\end{align}
or equivalently
\begin{align}
    P_+=& \frac{1}{8}\sum_{s= \pm 1} \sum_{s'= \pm 1} \int \int \int  d\hat{\mathbf{k}} d\bar{\hat{\mathbf{k}}}d\hat{\mathbf{k}'} f^{*}(\hat{\mathbf{k}}) f(\hat{\mathbf{k}'})\left[  \langle s'\vert s \rangle  e^{-i \Theta^{*}_{\hat{\mathbf{k}'}}} e^{-i s' \Omega^{*}_{\hat{\mathbf{k}'}}} e^{i \Theta_{\hat{\mathbf{k}}}} e^{i s\Omega_{\hat{\mathbf{k}}}}\delta_{\hat{\mathbf{k}'}, \bar{\hat{\mathbf{k}}}}  \delta_{\hat{\mathbf{k}}, \bar{\hat{\mathbf{k}}}}   \right.\nonumber \\
    & +   \langle s' \vert s \rangle e^{i\phi^{*}_{E}} e^{-i \Theta'^{*}_{\hat{\mathbf{k}'}}} e^{-i s'\Omega'^{*}_{\hat{\mathbf{k}'}}}e^{i \Theta_{\hat{\mathbf{k}}}} e^{i s\Omega_{\hat{\mathbf{k}}}}\delta_{\hat{\mathbf{k}}', \bar{\hat{\mathbf{k}}}} \delta_{\hat{\mathbf{k}}, \bar{\hat{\mathbf{k}}}} + \langle s'\vert s \rangle  e^{-i \Theta^{*}_{\hat{\mathbf{k}'}}} e^{-i s' \Omega^{*}_{\hat{\mathbf{k}'}}} e^{i \Theta'_{\hat{\mathbf{k}}}} e^{i s\Omega'_{\hat{\mathbf{k}}}}e^{-i\phi_{E}}\delta_{\hat{\mathbf{k}'}, \bar{\hat{\mathbf{k}}}} \delta_{\hat{\mathbf{k}}, \bar{\hat{\mathbf{k}}}}\nonumber \\  
    & + \left.
    \langle s'\vert s \rangle e^{i\phi^{*}_{E}} e^{-i '\Theta'^{*}_{\hat{\mathbf{k}'}}} e^{-i s \Omega'^{*}_{\hat{\mathbf{k}'}}}e^{i\phi_{E}} e^{i \Theta'_{\hat{\mathbf{k}}}} e^{i s\Omega'_{\hat{\mathbf{k}}}}\delta_{\hat{\mathbf{k}'}, \bar{\hat{\mathbf{k}}}}\delta_{\hat{\mathbf{k}}, \bar{\hat{\mathbf{k}}}}  \right].
\end{align}
 Integrating over $\bar{\hat{\mathbf{k}}}$ and $\hat{\mathbf{k}}'$ we have
\begin{align}\label{pf_dirac}
    P_+ =& \frac{1}{8}\sum_{s= \pm 1} \sum_{s'= \pm 1} \int  d\hat{\mathbf{k}} |f(\hat{\mathbf{k}})|^{2} \left[  \langle s'\vert s \rangle  e^{-i (\Theta^{*}_{\hat{\mathbf{k}}}- \Theta_{\hat{\mathbf{k}}} )} e^{-i s' \Omega^{*}_{\hat{\mathbf{k}}}}  e^{i s\Omega_{\hat{\mathbf{k}}}} \right.\nonumber 
     +   \langle s' \vert s \rangle e^{i\phi^{*}_{E}} e^{-i (\Theta'^{*}_{\hat{\mathbf{k}}} - \Theta_{\hat{\mathbf{k}}} )} e^{-i s'\Omega'^{*}_{\hat{\mathbf{k}}}} e^{i s\Omega_{\hat{\mathbf{k}}}} \nonumber\\
     &+ \langle s'\vert s \rangle  e^{-i (\Theta^{*}_{\hat{\mathbf{k}}}- \Theta'_{\hat{\mathbf{k}}})} e^{-i s' \Omega^{*}_{\hat{\mathbf{k}}}} e^{i s\Omega'_{\hat{\mathbf{k}}}}e^{-i\phi_{E}}  + \left.
    \langle s'\vert s \rangle  e^{-i( \Theta'^{*}_{\hat{\mathbf{k}}}- \Theta'_{\hat{\mathbf{k}}})} e^{-i s'\Omega'^{*}_{\hat{\mathbf{k}}}}e^{i s\Omega'_{\hat{\mathbf{k}}}}e^{i(\phi^{*}_{E}-\phi_{E})}  \right].
\end{align}
Now, an equally weighted state of helicity is given by
\begin{align}
    \frac{1}{\sqrt{2}}\sum_{s=\pm 1} e^{is\Omega}\vert s\rangle = \frac{1}{\sqrt{2}}\left( e^{i\Omega}\vert +1\rangle + e^{-i\Omega}\vert -1 \rangle\right).
\end{align}
The inner product between two such helicity states is 
\begin{eqnarray}
    \frac{1}{2}\sum_{s= \pm 1} \sum_{s'= \pm 1} \langle s'\vert s \rangle  e^{-i s'\Omega'^{*}_{\hat{\mathbf{k}}}}e^{i s\Omega'_{\hat{\mathbf{k}}}}
    &=& \frac{1}{2}\left( \langle +1 \vert e^{-i\Omega'^{*}} + \langle -1 \vert e^{i\Omega'^{*}}\right)\left( e^{i\Omega'}\vert +1\rangle + e^{-i\Omega'}\vert -1 \rangle\right), 
    \nonumber \\
    &=& \frac{1}{2}\left(e^{-i(\Omega'^{*} - \Omega')} + e^{i(\Omega'^{*} - \Omega')}\right), 
    \nonumber \\
    &=& \cos\left(\Omega'^{*}_{\hat{\mathbf{k}}} - \Omega'_{\hat{\mathbf{k}}}\right).
\end{eqnarray}
Using this result, the  probability of detection \eqref{pf_dirac} becomes 
\begin{align}
    P_+ =& \frac{1}{4} \int  d\hat{\mathbf{k}} |f(\hat{\mathbf{k}})|^{2} \left[ e^{-i (\Theta^{*}_{\hat{\mathbf{k}}}- \Theta_{\hat{\mathbf{k}}} )}\cos\left(\Omega^{*}_{\hat{\mathbf{k}}} - \Omega_{\hat{\mathbf{k}}}\right) \right.\nonumber 
     +    e^{i\phi^{*}_{E}} e^{-i (\Theta'^{*}_{\hat{\mathbf{k}}} - \Theta_{\hat{\mathbf{k}}} )} \cos\left(\Omega'^{*}_{\hat{\mathbf{k}}} - \Omega_{\hat{\mathbf{k}}}\right) \nonumber\\
     &+ e^{-i (\Theta^{*}_{\hat{\mathbf{k}}}- \Theta'_{\hat{\mathbf{k}}})}e^{-i\phi_{E}}\cos\left(\Omega^{*}_{\hat{\mathbf{k}}} - \Omega'_{\hat{\mathbf{k}}}\right)  + \left.
     e^{-i( \Theta'^{*}_{\hat{\mathbf{k}}}- \Theta'_{\hat{\mathbf{k}}})}e^{i(\phi^{*}_{E}-\phi_{E})}\cos\left(\Omega'^{*}_{\hat{\mathbf{k}}} - \Omega'_{\hat{\mathbf{k}}}\right)  \right].
\end{align}
Since the phase from the gravitational time delay and the Wigner phase are real, the probability becomes then
\begin{align}
    P_+ =& \frac{1}{4} \int  d\hat{\mathbf{k}} |f(\hat{\mathbf{k}})|^{2} \left[ 1 + e^{i(\phi^{*}_{E}-\phi_{E})} \right.\nonumber \\
     &+\left. e^{i\phi^{*}_{E}} e^{-i (\Theta'_{\hat{\mathbf{k}}} - \Theta_{\hat{\mathbf{k}}} )} \cos\left(\Omega'_{\hat{\mathbf{k}}} - \Omega_{\hat{\mathbf{k}}}\right)+ e^{-i (\Theta_{\hat{\mathbf{k}}}- \Theta'_{\hat{\mathbf{k}}})}e^{-i\phi_{E}}\cos\left(\Omega_{\hat{\mathbf{k}}} - \Omega'_{\hat{\mathbf{k}}}\right) \right]
\end{align}
or equivalently
\begin{align}
    P_+ =& \int  d\hat{\mathbf{k}} |f(\hat{\mathbf{k}})|^{2} \left[ \frac{1}{2}  +\frac{1}{4}\cos\left(\Omega'_{\hat{\mathbf{k}}} - \Omega_{\hat{\mathbf{k}}}\right)\left(e^{-i (\Theta'_{\hat{\mathbf{k}}} - \Theta_{\hat{\mathbf{k}}} +\phi_{E})} + e^{i(\Theta'_{\hat{\mathbf{k}}}-\Theta_{\hat{\mathbf{k}}} +\phi_{E})}\right)\right]. \nonumber
\end{align}
Identifying $\Delta\vartheta = \Omega'_{\hat{\mathbf{k}}} - \Omega_{\hat{\mathbf{k}}}$ and $\Delta\vartheta^{\tau} = \Theta'_{\hat{\mathbf{k}}} - \Theta_{\hat{\mathbf{k}}}$ we finally obtain for the probability of detection the following expression 
\begin{align}\label{ProbMZa}
    P_{\pm} = \frac{1}{2}\left[1 \pm \int d\hat{\mathbf{k}} \vert f(\hat{\mathbf{k}}) \vert^2 \cos \left( \Delta\vartheta \right)\cos\left( \Delta\vartheta^{\tau} +\phi_{E}\right) \right].
\end{align}


\section{General treatment of null vector and tetrad basis}
\label{Appendix:General_nullgeodesics}

We express the null vector $k^{\mu}$ as 
\begin{equation}
    k^{\mu} = \left( c\frac{dt}{d\lambda} \, , \, \frac{dr}{d\lambda},  \, \frac{d\theta}{d\lambda},  \, \frac{d\varphi}{d\lambda}\right),
\end{equation}
where $\lambda$ is an affine parameter along the trajectory. For a photon following an arbitrary geodesic in the Kerr spacetime, the null vector can be expressed as \cite{Felice_Bini_2010,MTW1973}
\begin{align}\label{null_complete}
   c\frac{dt}{d\lambda} =& \, \frac{1}{\rho^2}\left[-a\left(a \mathcal{T} \sin^{2}\theta - \mathcal{R} \right) + \frac{\left( r^{2} + a^{2}\right)}{\Delta} P \right],   \\ 
  \frac{dr}{d\lambda} =& \,  \epsilon_r \frac{1}{\rho^2} \sqrt{R}, \\ 
   \frac{d\theta}{d\lambda} =& \,  \epsilon_{\theta}\frac{1}{\rho^2}\sqrt{\Theta} , \\ 
    \frac{d\varphi}{d\lambda} =& \,  -\frac{1}{\rho^2}\left[ \left(a \mathcal{T} - \frac{\mathcal{R}}{\sin^{2}\theta }\right) + \frac{a}{\Delta} P\right],
\end{align}
where $\epsilon_r$ and $\epsilon_{\theta}$ are sign indicators, i.e., $\epsilon_r =\pm1$, $\epsilon_\theta =\pm 1$, and
\begin{align}
    P =& \, \mathcal{T}\left( r^{2} + a^{2}\right) - a \mathcal{R}, \\
    R =& \, P^{2} - \Delta\left[ \mathcal{K} + \left( \mathcal{R}- a\mathcal{T}\sin^{2}\theta\right)^{2}\right],\\
    \Theta =& \mathcal{K} - \cos^{2}\theta \left[ -a^{2} \mathcal{T}^{2} + \frac{\mathcal{R}^{2}}{\sin^{2}\theta}\right].
   \end{align}
Here, the quantities $\mathcal{R}$, $\mathcal{T}$ and $\mathcal{K}$ are constants of the motion representing the azimuthal angular momentum, the total energy and the separation constant of the Hamilton-Jacobi equation~\cite{Felice_Bini_2010}, respectively.  Introducing the notation
\begin{equation}
    b= \frac{\mathcal{R}}{\mathcal{T}} ,\qquad q^{2} =\frac{\mathcal{K}}{\mathcal{T}^2},
\end{equation}
the equations above become
\begin{eqnarray}
P &=& \mathcal{T}P'=\mathcal{T} \left[(r^{2} +a^{2}) - ab\right], \\
R &=& \mathcal{T}^{2}R' = \mathcal{T}^{2}\left[(r^{2} + a^{2}-ab)^{2} - \Delta(q^{2} + (b-a\sin^{2}\theta)^{2}) \right],\\
\Theta &=& \mathcal{T}^{2}\Theta' =\mathcal{T}^{2}\left[q^{2} - \cos^{2}\theta\left(-a^{2} + \frac{b^{2}}{\sin^{2} \theta} \right)  \right].
\end{eqnarray}
Thus, the components of the null vector are
\begin{align}\label{null_complete_1}
   c\frac{dt}{d\lambda} =& \, \frac{\mathcal{T}}{\rho^2}\left[-a\left(a  \sin^{2}\theta - b \right) + \frac{\left( r^{2} + a^{2}\right)}{\Delta} (r^{2} +a^{2} -ab) \right],   \\ 
  \frac{dr}{d\lambda} =& \,  \epsilon_r \frac{\mathcal{T}}{\rho^2} \sqrt{R'}, \\ 
   \frac{d\theta}{d\lambda} =& \,  \epsilon_{\theta}\frac{\mathcal{T}}{\rho^2}\sqrt{\Theta'} , \\ 
    \frac{d\varphi}{d\lambda} =& \,  -\frac{\mathcal{T}}{\rho^2}\left[ \left(a  - \frac{b}{\sin^{2}\theta }\right) + \frac{a}{\Delta} (r^{2} + a^{2} -ab)\right].
\end{align}
Now, considering the integral curves $u^{\mu} = k^{\mu}/\mathcal{T}$, that is, $\lambda \rightarrow \mathcal{T}\lambda'$ and $\lambda' \rightarrow \lambda$, the previous equations become
\begin{align}\label{null_complete_final}
   c\frac{dt}{d\lambda} =& \, \frac{1}{\rho^{2}}\left[-a\left(a  \sin^{2}\theta - b \right) + \frac{\left( r^{2} + a^{2}\right)}{\Delta} (r^{2} +a^{2} -ab) \right],   \\ 
  \frac{dr}{d\lambda} =& \,  \epsilon_r \frac{1}{\rho^2} \sqrt{R'}, \\ 
   \frac{d\theta}{d\lambda} =& \,  \epsilon_{\theta}\frac{1}{\rho^2}\sqrt{\Theta'} , \\ 
    \frac{d\varphi}{d\lambda} =& \,  -\frac{1}{\rho^2}\left[ \left(a  - \frac{b}{\sin^{2}\theta }\right) + \frac{a}{\Delta} (r^{2} + a^{2} -ab)\right].
\end{align}

The observer that prepares and measures the quantum states in our scheme is a zero-angular-momentum observer (ZAMO). In this case, the time-like $e^{\mu}_{\hat{0}}$ component is given by
\begin{equation}
    e^{\mu}_{\hat{0}} := \frac{u^{\mu}_{\rm ZAMO}}{c}.
\end{equation}
Now we can make a general expression for the basis of tetrads under the condition that $u^{\mu} \propto e^{\mu}_{0} + e^{\mu}_{j}$ for $j = 1,2,3$ and considering the normalization $e^{\hat{0}}_{\mu}dx^{\mu}/d\lambda=1$, that is, $ u^{\mu} = u^{\hat{0}}\bar{u}^{\mu} $, where
\begin{equation}
    u^{\hat{0}} = \left(g_{00} - \frac{g_{03}^2}{g_{33}}\right)\frac{Z}{c} u^{0},
\end{equation}
with $Z$ the normalization constant.
The tetrad basis is then the following
\begin{align}
e^{\mu}_{\hat{0}}=& \left(e^{0}_{\hat{0}},0,0,e^{3}_{\hat{0}}\right), \\
\nonumber \\
e^{\mu}_{\hat{1}}=&\left(\frac{(e^{0}_{\hat{0}}g_{03}+e^{3}_{\hat{0}}g_{33}) \sqrt{g_{11} (\bar{u}^{1})^2+g_{22} (\bar{u}^{2})^2}}{\sqrt{\left(g_{00} g_{33}-g_{03}^2\right)((e^{3}_{\hat{0}})^2(A))}}, \right. \nonumber \\
& \frac{\bar{u}^{1}( e^{3}_{\hat{0}}\bar{u}^{0} -  e^{0}_{\hat{0}}\bar{u}^{3}) \left(g_{03}^2-g_{00} g_{33}\right)}{\sqrt{g_{11} (\bar{u}^{1})^2 + g_{22}(\bar{u}^{2})^2} \sqrt{\left(g_{00} g_{33}-g_{03}^2\right) ((e^{3}_{\hat{0}})^2(A))}},  \nonumber \\
&\frac{\bar{u}^{2}( e^{3}_{\hat{0}}\bar{u}^{0} -  e^{0}_{\hat{0}}\bar{u}^{3}) \left(g_{03}^2-g_{00} g_{33}\right)}{\sqrt{g_{11} (\bar{u}^{1})^2 + g_{22}(\bar{u}^{2})^2} \sqrt{\left(g_{00} g_{33}-g_{03}^2\right) ((e^{3}_{\hat{0}})^2(A))}},  \nonumber \\
&\left.-\frac{(e^{0}_{\hat{0}}g_{00}+e^{3}_{\hat{0}}g_{03}) \sqrt{g_{11} (\bar{u}^{1})^2+g_{22} (\bar{u}^{2})^2}}{\sqrt{\left(g_{00} g_{33}-g_{03}^2\right)((e^{3}_{\hat{0}})^2(A))}} \right),
\\
e^{\mu}_{\hat{2}}=& \left(0,\frac{\bar{u}^2\sqrt{g_{22}}}{\sqrt{-g_{11} \left(g_{11} (\bar{u}^{1})^2 + g_{22} (\bar{u}^{2})^2\right)}},-\frac{g_{11} \bar{u}^{1}}{\sqrt{g_{22}} \sqrt{-g_{11} \left(g_{11} (\bar{u}^{1})^2 + g_{22} (\bar{u}^{2})^2\right)}},0\right),\\ 
\nonumber\\
e^{\mu}_{\hat{3}}=& \left(\bar{u}^{0}-e^{0}_{\hat{0}},\bar{u}^{1},\bar{u}^{2},\bar{u}^{3}-e^{3}_{\hat{0}}\right),
\end{align}

where
\begin{align}
    A=&\, g_{03}^2(\bar{u}^0)^2-g_{33}(g_{00}(\bar{u}^0)^2 + g_{11}(\bar{u}^1)^2 +g_{22}(\bar{u}^2)^2) -2e^{0}_{\hat{0}}e^{3}_{\hat{0}}(g_{03}(g_{11}(\bar{u}^1)^2 + g_{22}(\bar{u}^2)^2 + g_{03} \bar{u}^0\bar{u}^3) -g_{00}g_{33}\bar{u}^0\bar{u}^3) \nonumber\\
    & - (e^{0}_{\hat{0}})^2(-g_{03}^2(\bar{u}^3)^2 + g_{00}(g_{11}(\bar{u}^1)^2 + g_{22}(\bar{u}^2)^2 +g_{33}(\bar{u}^3)^2)).
\end{align}

En la forma en que esta escrito, se puede considerar el caso de observador estatico con $e^{3}_{\hat{0}}=0$ y $e^{0}_{\hat{0}}= 1/\sqrt{g_{tt}} = u^{0}_{\rm Static}/c$, es decir, $e^{\mu}_{\hat{0}}= u^{\mu}_{\rm Static}/c$

\section{Error analysis}\label{Appendix:Error analysis}

To obtain an expression $\delta a$ for the error in estimating the specific angular momentum $a$, we use error propagation assuming that the only source of error is the measurement of the detection probability $P_+$. For this, we define the following quantities $\Lambda = c^2 |h||l|$, $\Gamma = m^2 \cos^2(\theta)$, and $\Omega = 2\sigma^2 + \omega^2$. In addition, we define the following:
\begin{equation}
    \varsigma = \sqrt{\Lambda^2 \Gamma^2 - (P_{+} - 1)r_2^{10}\Omega}.
\end{equation}

The expression for $\delta a$, given an error $\delta p$ in the detection probability $P_+$, is 
\begin{equation}
    \delta a = \frac{\Omega \, r_2^{10} \, \delta p}{4 \varsigma \left( \Lambda \Gamma - \varsigma \right)}.
    \label{eq:sigma_a_canonical}
\end{equation}

 \section{ZAMO observers}

 The four velocity of the ZAMO observer is given by 
 \begin{equation}
     u^{\mu}_{\rm ZAMO} = Z(1,0,0,\omega),
 \end{equation}
 where $\omega= -(g_{03}/g_{33})$  and
 \begin{equation}
     Z= \frac{c}{\sqrt{g_{00} +2\omega g_{03} + \omega^2 g_{33}}}.
 \end{equation}

\end{widetext}

\bibliography{references} 

\end{document}